 \PassOptionsToPackage{table}{xcolor}
\documentclass[acmsmall,nonacm]{acmart}

\setcopyright{none}

 \title[]{SafeNom: Data-Aware Microservice Policies}
 \author{Karuna Grewal}
\email{kgrewal@cs.cornell.edu}
\affiliation{%
  \institution{Cornell University}
  \country{USA}
}
\author{P. Brighten Godfrey}
\affiliation{%
  \institution{University of Illinois
Urbana-Champaign}
  \country{USA}
}
\email{pbg@illinois.edu}
\author{Justin Hsu}
\affiliation{%
  \institution{Cornell University}
  \country{USA}
}
\email{justin@cs.cornell.edu}
 \setcopyright{none}

 \usepackage[normalem]{ulem} % Enables \sout
 \usepackage[safe]{tipa}
 \usepackage{forest}
 \definecolor{safenomLogin}{RGB}{38, 99, 190}
\definecolor{safenomMFA}{RGB}{210, 115, 20}

 \usepackage{booktabs}   %% For formal tables:
 \usepackage{subcaption} %% For complex figures with subfigures/subcaptions
 \newcommand{\node}[0]{\textsf{node}}
 \usepackage{tcolorbox}
 \usepackage{syntax}
 \usepackage{listings}
 \usepackage{comment}
 \makeatletter
 \@addtoreset{equation}{enumi}
 \makeatother
 \usepackage{mathpartir}
 
 \makeatletter
 \let\old@lstKV@SwitchCases\lstKV@SwitchCases
 \def\lstKV@SwitchCases#1#2#3{}
 \makeatother
 \usepackage{lstlinebgrd}
 \makeatletter
 \let\lstKV@SwitchCases\old@lstKV@SwitchCases
 
 \lst@Key{numbers}{none}{%
     \def\lst@PlaceNumber{\lst@linebgrd}%
     \lstKV@SwitchCases{#1}%
     {none:\\%
      left:\def\lst@PlaceNumber{\llap{\normalfont
                 \lst@numberstyle{\thelstnumber}\kern\lst@numbersep}\lst@linebgrd}\\%
      right:\def\lst@PlaceNumber{\rlap{\normalfont
                 \kern\linewidth \kern\lst@numbersep
                 \lst@numberstyle{\thelstnumber}}\lst@linebgrd}%
     }{\PackageError{Listings}{Numbers #1 unknown}\@ehc}}
 \makeatother
 \usepackage{lstlinebgrd}
 \usepackage{hyperref}
 \usepackage{cleveref}
 \Crefname{figure}{Fig.}{Figs.}
 
 \newif\ifappendix
 \appendixfalse
 \newcommand{\appref}[1]{%
  \ifappendix
    \S\ref{#1}
  \else
  \fi
}
 \makeatletter
\def\@copyrightpermission{}
\def\@acmDOI{}
\def\@acmPrice{}
\def\@acmISBN{}
\makeatother
 \usepackage{titletoc}
 \usepackage{multirow}
 \usepackage{makecell}
 \makeatletter
 \newcommand{\bigplus}{%
   \DOTSB\mathop{\mathpalette\mattos@bigplus\relax}\slimits@
 }
 \newcommand\mattos@bigplus[2]{%
   \vcenter{\hbox{%
     \sbox\z@{$#1\sum$}%
     \resizebox{!}{0.9\dimexpr\ht\z@+\dp\z@}{\raisebox{\depth}{$\m@th#1+$}}%
   }}%
   \vphantom{\sum}%
 }
 \usepackage{custommacro}
\definecolor{thmBg}{gray}{0.93}
\definecolor{thmBorder}{gray}{0.55}
\tcbuselibrary{breakable}
\newcounter{theorem}

\newenvironment{theorem}{%
  \refstepcounter{theorem}%
  \begin{tcolorbox}[colback=thmBg, colframe=thmBorder, boxrule=0.5pt,
    arc=3pt, left=8pt, right=8pt, top=6pt, bottom=6pt,
    before skip=10pt, after skip=10pt, breakable]
  \normalfont\noindent\textbf{Theorem \thetheorem.}\par\vspace{2pt}\noindent\ignorespaces
}{%
  \end{tcolorbox}
}
\begin{document}
\begin{abstract}
  Many cloud-based applications are organized as loosely coupled microservices,
  where invoking a service's API triggers a cascade of APIs across many services
  and leads to inter-service exchange of API parameters and output responses. 
Current tools for monitoring microservice safety properties have limited
expressiveness for properties that describe the flow of data through API calls. 

To this end, we present \safenom, a specification and monitoring framework for
microservices based on nominal languages. \safenom policies can express both the
desired order of API calls and how the data carried in requests and
responses should or should not flow between the APIs.
Policies are enforced using a nominal automaton-based distributed runtime
monitor which can be applied in a blackbox and non-invasive manner, without
access to the service implementation and without making changes to the service
implementation.
Our experiments show that our monitor can efficiently enforce rich data-aware
properties while incurring minimal latency overhead, on the order of a few
milliseconds.
\end{abstract}
\maketitle   
\section{Introduction}
A popular paradigm for implementing a modern-day cloud application is to decompose it into loosely
coupled self-contained microservices that expose their functionality through APIs, which can be accessed using protocols like HTTP or gRPC.
In such an application, a single user request can trigger a cascade of downstream API calls and responses,  resulting in a tree-shaped execution. Each call or response may carry data as parameters or results. For instance, these data include region identifiers, access tokens, confidential data, encryption keys, or trace IDs.  Correct handling of this data is crucial  to application correctness, and verifying that it is handled properly can  provide assurance to various teams: compliance, development, and deployment.

Since data moves between services, handling it correctly is an application-wide concern rather than the responsibility of any single service. A service may correctly process its local input and invoke the correct sequence of APIs, yet the overall execution can still be unsafe if a later service receives or uses the wrong value. For instance, a database write may occur after an encryption API to ensure compliant storage of personally identifiable information, but the execution may be unsafe if the value written to the database is the original raw data rather than the encrypted result. Thus, many policies must specify not only which services should be called and in what order, but also how the values carried by those calls relate to one another. Violations of such policies can cause data leaks, broken auditability, regulatory penalties, and loss of client trust, even when each individual service appears to satisfy its local API contract. 

Enforcement of such data-aware policies is complicated by  the distributed design and deployment of microservice applications. The services in a microservice application are often not controlled by a single team. They may be  developed by independent teams that do not have access to each other's implementations, deployed by another team, and audited by a separate security or compliance team. 
Consequently, the team tasked with certifying an application may not have access to all service implementations and even when it does, requiring global changes to all services may be impractical. Thus, an  enforcement mechanism should treat the application as a blackbox. Inter-service communication provides a natural boundary for teams to specify and enforce fine-grained aspects of the application behavior without peeking into service code.

\newpar{Existing work and limitations}
Existing deployment infrastructure and runtime monitors for microservices can enforce constraints on which APIs may communicate and in what order, but they remain unable to enforce properties that relate data values across an execution. For example, Kubernetes, a container network orchestration tool, allows a user to coarsely
restrict or allow all access to services. Meanwhile,  \emph{service
mesh}~\cite{servicemesh}, a recent networking layer,  offers finer-grained enforcement of inter-service communication policies, but its policies are \textit{single} hop, meaning they can only specify whether a pair of endpoints is allowed to
communicate through API calls. 

Recent runtime monitoring frameworks 
\cite{grewalhotnets,safetree} for microservice applications move beyond single-hop checks by enforcing multi-hop policies over the execution's tree structure. The work by \citet{grewalhotnets} introduces regular expression-style policies over the order of API calls, while SafeTree \cite{safetree} adds declarative constructs for specifying valid parent-child structure in the API call tree. Both systems are implemented as distributed monitors at the service mesh networking layer, using  finite-state and visibly pushdown automata, respectively. This mirrors the distributed deployment structure of a microservice application. Each service's incoming/outgoing traffic is locally monitored without a centralized monitor. The monitors are also blackbox and non-invasive, meaning they require no access to, or modifications of, the service implementation. However, neither framework can enforce properties that depend on the values carried by API calls or responses. This limitation rules out many natural microservice properties, like:
\begin{enumerate}
  \item A compliance team may want the raw data to be encrypted before the encrypted result (\textit{different} from the raw data) is stored in a database.
  \item A development team may want that in a multi-party co-signature protocol,
    within a session the APIs called by each party use the \textit{same} key,
    while different sessions may use distinct keys.
  \item A deployment team may want all requests in an execution trace to carry  the \textit{same} trace ID parameter.
\end{enumerate}

These examples share a common form: they require reasoning about equality and
inequality patterns among data carried by API calls/responses, not just the
order of API calls. We call such properties \textit{data-aware}, and in this
work we consider how to specify and monitor these properties.

We target a setting where teams that specify properties do not have access to
the service implementation, and must treat the application as a blackbox. Thus,
we consider the following question:
\textit{``How can we specify and enforce data-aware microservice policies in a
blackbox and non-invasive manner, \textit{i.e.,} without access to the service
implementation and without invasive changes to the service code?''} Addressing this question requires solving the following three challenges. 

\newpar{Challenge 1: Policies should depend on (in)equality patterns and not concrete values}
A common theme across our data-aware policies is that they specify an equality or inequality relation on the values of certain parameters across
several API calls, but the concrete value of the parameter in the trace is
insignificant.  For instance, the trace-ID policy (3) above should accept any execution in which all API calls share the same trace ID, whether that ID is $10$ or $20$. Similarly, the (1) encryption policy above should accept any execution based on whether certain data values are (in)equal. Thus, the policies should be invariant under consistent renaming of values: two executions that differ only by replacing concrete values while preserving the same equality and inequality patterns should either both satisfy the property or violate it. This rules out treating  values as elements of a fixed finite alphabet because the policy should not enumerate all possible values that may appear at runtime.  Therefore, \textit{a suitable specification language should treat traces that preserve the (in)equality patterns modulo the concrete parameter/result values.}

\newpar{Challenge 2: (In)equality constraints are scoped.}
(In)equality relationships in microservice executions are not always global. Some values must remain consistent across an entire execution, while others may be relevant only within a smaller session, or a protocol phase, or a child call. For instance, the (3) trace ID policy above imposes a constraint that the trace ID remain fixed across the \textit{entire} execution. In contrast, the (2)  co-signature policy  establishes the key equality constraint per-child session. After a session returns, the next session's keys can be different. The policy author should have the flexibility to define the beginning and end of the scope of a  data-dependency obligation. Thus, scoping is a semantic requirement of the policy language.

\newpar{Challenge 3: Efficient enforcement over a large value domain.}
Parameter and return values in a microservice trace range over a large
domain---a monitor cannot use one automaton state per possible value. Moreover,
in our blackbox microservice setting, the monitor must simulate an automaton in
a distributed manner and carry the relevant monitoring metadata in HTTP headers.
To fit this monitoring metadata into the limited header space,  the monitor
should track only values that are currently in scope and needed for a later
comparison, while dropping values when they go out of scope. Thus, the
underlying automaton needs a disciplined strategy to decide \textit{when} to start tracking a   value, \textit{how long} to retain it for later comparisons, and \textit{when} to  discard it once policy checking no longer requires it.

\subsection*{Our approach}
To address these challenges, we propose \safenom, a  specification language and monitoring
framework for data-aware microservice policies.

\newpar{Nominal words as the semantic foundation.}
We use nominal words and nominal languages to make invariance under value renaming an intrinsic property of the policy semantics. 
Intuitively, a nominal word is a sequence of ordinary symbols, like API call and
return events, augmented with special symbols called \textit{names}, which
correspond to data values in our setting. Then, a nominal language is a set of nominal words such that freely replacing a name
keeps the word in the language.
This makes nominal languages a suitable interpretation for our data-aware
properties, which should only depend on repetition patterns of (in)equal values
rather than concrete values.

\newpar{Lazy binders for scoping}
We develop the \safenom policy language for expressing \textit{scoped data-dependencies} over microservice traces. Our policies can declare when the tracking of  a value begins and ends for a data-dependency constraint. Our starting point is the 
nominal regular expressions with binders (NREs) introduced by \citet{kurznomautomata}, whose binders provide a declarative mechanism to introduce names for values and delimit the part of the trace over which they can be referenced. 

However, Kurz et al.'s NRE \textit{eager} binders are too restrictive for
modeling microservice traces because they require the value associated with a
data-dependency constraint to appear exactly where the constraint's scope
begins. However, in microservice executions, the scope of a data-dependency may begin before the relevant value appears in an API event.  \safenom addresses this mismatch with \textit{lazy binders}. A lazy binder opens the scope of a name without immediately assigning it a value. The name is initialized at the point of its first observation. Once initialized, it can be used for future comparison. Thus, lazy binders let the policy specify the scope of a constraint while deferring the value's binding to the point where it is observed in the trace. 
\safenom further extends NREs with semantics for inequality checks because NREs
are designed around equality checks on names, whereas many interesting
properties in the microservice setting also require checking for
\emph{inequality} between certain data.

\newpar{Nominal automaton-based monitoring}
To enforce \safenom policies, we compile them into a nominal automaton model
inspired by \citet{kurznomautomata}, extended to support SafeNom's lazy binding
and inequality semantics. Building on SafeTree's automaton-based distributed
monitoring setup, we implement the compiled nominal automaton as a runtime
monitor that simulates the automaton's transitions on
the execution trace of HTTP messages in a microservice application. Since a nominal automaton is finite-state regardless of the size of the value domain, and it tracks only the names currently in scope, the monitoring metadata stays within the HTTP header budget.
By implementing our monitor atop the \textit{service mesh} \cite{istio}
networking layer, it runs outside the service containers and does not
require any visibility or invasive changes to the service implementation. 

\paragraph*{Contributions.}
We make the following technical contributions:
\begin{enumerate} 
\item an extension of nominal words and nominal regular expressions supporting
  \textit{lazy binding} (\cref{sec:extension}), 
\item the syntax and semantics of our \safenom language for specifying
  data-aware microservice policies (\cref{sec:safenom}),
\item case studies that demonstrate SafeNom's expressiveness (\cref{sec:casestudies}),
\item a method to soundly compile \safenom policies to a novel notion of nominal
  automata (\cref{sec:enforcement}), and
\item a \textit{service mesh-based} implementation of a runtime monitor
  for \safenom policies (\cref{sec:impl}) and an evaluation of its performance
  (\cref{sec:eval}).
\end{enumerate}

\section{A Tour of \safenom Policies}\label{sec:tour}
This section motivates \safenom through a multi-factor authentication application. 
The goal is to illustrate why data-aware policies require more than checking
which APIs are called, or in which order they are called. Many safety properties
for microservice applications are about the \emph{flow of values} through API
parameters and responses: a value received by one service should be forwarded
unchanged to another service, a transformed value should be used instead of the
raw value, or two values should be provably different. \safenom provides a
policy language and runtime monitor for such properties.

\subsection{Example: login with multi-factor authentication}
Consider an authentication service that implements login with multi-factor authentication (MFA). The application exposes a \login API that takes in a password as its parameter. After checking the input password, the \login service calls  the \mfa API to perform one round of multi-factor authentication. Each \mfa call carries a  challenge code as its input. This is the code that will be shown to the user through an authenticator application. The corresponding \mfa response carries the code confirmed by the user. In high-risk login settings, the application may issue multiple MFA challenges for additional verification. For simplicity, we assume that the application uses two MFA challenges.

\newpar{Safety property}
A security team may require the following data-flow property:
\begin{enumerate}
\item \login execution should include two calls to \mfa APIs.  
\item Each MFA response must return the same challenge code that was passed to its corresponding MFA call.
\item The MFA challenge code must be distinct from the enclosing login password. This is because the password and the one-time challenge codes serve different roles in the login process.  
\end{enumerate}
This property constrains both the order of API calls and returns and the values they carry.

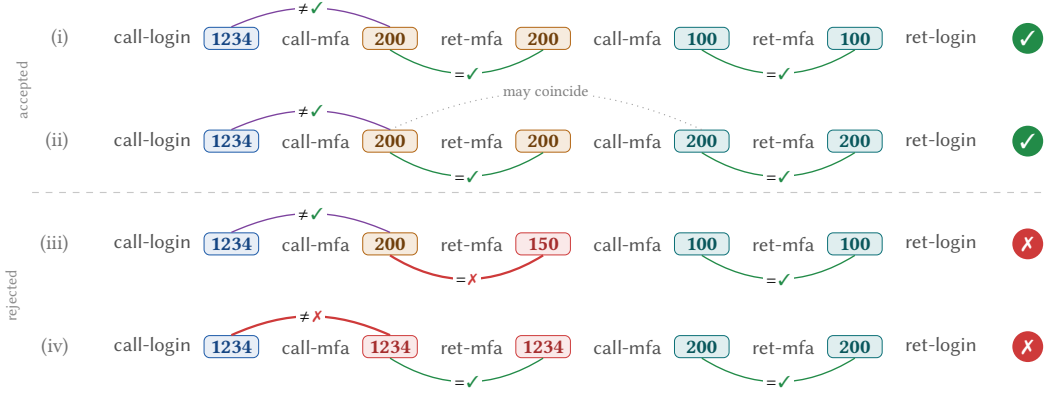
\begin{figure}[]
\centering
\snfit{%
\begin{tikzpicture}
\node[ev] (r1e1) at (0,0) {\clogin};
\node[pw,right=\evgap of r1e1]  (r1v1) {1234};
\node[ev,right=\grpgap of r1v1] (r1e2) {\cmfa};
\node[mf,right=\evgap of r1e2]  (r1v2) {200};
\node[ev,right=\grpgap of r1v2] (r1e3) {\rmfa};
\node[mf,right=\evgap of r1e3]  (r1v3) {200};
\node[ev,right=\grpgap of r1v3] (r1e4) {\cmfa};
\node[mf2,right=\evgap of r1e4]  (r1v4) {100};
\node[ev,right=\grpgap of r1v4] (r1e5) {\rmfa};
\node[mf2,right=\evgap of r1e5]  (r1v5) {100};
\node[ev,right=\grpgap of r1v5] (r1e6) {\rlogin};
\node[okbadge,right=12pt of r1e6] (r1b) {\ding{51}};
\node[rowlbl,left=13pt of r1e1]   (r1l) {(i)};
\draw[eqarc]  (r1v2.south) to[bend right=32] node[arclbl]{$=$\,\okmark} (r1v3.south);
\draw[eqarc]  (r1v4.south) to[bend right=32] node[arclbl]{$=$\,\okmark} (r1v5.south);
\draw[neqarc] (r1v1.north) to[bend left=24]  node[arclbl]{$\neq$\,\okmark} (r1v2.north);

\node[ev] (r2e1) at (0,-1.6) {\clogin};
\node[pw,right=\evgap of r2e1]  (r2v1) {1234};
\node[ev,right=\grpgap of r2v1] (r2e2) {\cmfa};
\node[mf,right=\evgap of r2e2]  (r2v2) {200};
\node[ev,right=\grpgap of r2v2] (r2e3) {\rmfa};
\node[mf,right=\evgap of r2e3]  (r2v3) {200};
\node[ev,right=\grpgap of r2v3] (r2e4) {\cmfa};
\node[mf2,right=\evgap of r2e4]  (r2v4) {200};
\node[ev,right=\grpgap of r2v4] (r2e5) {\rmfa};
\node[mf2,right=\evgap of r2e5]  (r2v5) {200};
\node[ev,right=\grpgap of r2v5] (r2e6) {\rlogin};
\node[okbadge,right=12pt of r2e6] (r2b) {\ding{51}};
\node[rowlbl,left=13pt of r2e1]   (r2l) {(ii)};
\draw[eqarc]  (r2v2.south) to[bend right=32] node[arclbl]{$=$\,\okmark} (r2v3.south);
\draw[eqarc]  (r2v4.south) to[bend right=32] node[arclbl]{$=$\,\okmark} (r2v5.south);
\draw[neqarc] (r2v1.north) to[bend left=24]  node[arclbl]{$\neq$\,\okmark} (r2v2.north);
\draw[neutarc] (r2v2.north) to[bend left=24]
      node[arclbl,text=black!50]{may coincide} (r2v4.north);

\node[ev] (r3e1) at (0,-3.2) {\clogin};
\node[pw,right=\evgap of r3e1]  (r3v1) {1234};
\node[ev,right=\grpgap of r3v1] (r3e2) {\cmfa};
\node[mf,right=\evgap of r3e2]  (r3v2) {200};
\node[ev,right=\grpgap of r3v2] (r3e3) {\rmfa};
\node[bad,right=\evgap of r3e3] (r3v3) {150};
\node[ev,right=\grpgap of r3v3] (r3e4) {\cmfa};
\node[mf2,right=\evgap of r3e4]  (r3v4) {100};
\node[ev,right=\grpgap of r3v4] (r3e5) {\rmfa};
\node[mf2,right=\evgap of r3e5]  (r3v5) {100};
\node[ev,right=\grpgap of r3v5] (r3e6) {\rlogin};
\node[xbadge,right=12pt of r3e6] (r3b) {\ding{55}};
\node[rowlbl,left=13pt of r3e1]  (r3l) {(iii)};
\draw[vioarc] (r3v2.south) to[bend right=32] node[arclblv]{$=$\,\nomark} (r3v3.south);
\draw[eqarc]  (r3v4.south) to[bend right=32] node[arclbl]{$=$\,\okmark} (r3v5.south);
\draw[neqarc] (r3v1.north) to[bend left=24]  node[arclbl]{$\neq$\,\okmark} (r3v2.north);
 
\node[ev] (r4e1) at (0,-4.8) {\clogin};
\node[pw,right=\evgap of r4e1]  (r4v1) {1234};
\node[ev,right=\grpgap of r4v1] (r4e2) {\cmfa};
\node[bad,right=\evgap of r4e2] (r4v2) {1234};
\node[ev,right=\grpgap of r4v2] (r4e3) {\rmfa};
\node[bad,right=\evgap of r4e3] (r4v3) {1234};
\node[ev,right=\grpgap of r4v3] (r4e4) {\cmfa};
\node[mf2,right=\evgap of r4e4]  (r4v4) {200};
\node[ev,right=\grpgap of r4v4] (r4e5) {\rmfa};
\node[mf2,right=\evgap of r4e5]  (r4v5) {200};
\node[ev,right=\grpgap of r4v5] (r4e6) {\rlogin};
\node[xbadge,right=12pt of r4e6] (r4b) {\ding{55}};
\node[rowlbl,left=13pt of r4e1]  (r4l) {(iv)};
\draw[eqarc]  (r4v2.south) to[bend right=32] node[arclbl]{$=$\,\okmark} (r4v3.south);
\draw[eqarc]  (r4v4.south) to[bend right=32] node[arclbl]{$=$\,\okmark} (r4v5.south);
\draw[vioarc] (r4v1.north) to[bend left=24]  node[arclblv]{$\neq$\,\nomark} (r4v2.north);
 
\draw[divider] ([yshift=8mm]r3l.west) -- ([yshift=8mm]r3b.east);
\node[grouptag] at ([xshift=-3mm,yshift=-8mm]r1l.west) {accepted};
\node[grouptag] at ([xshift=-3mm,yshift=-8mm]r3l.west) {rejected};
\end{tikzpicture}
}
\caption{Example \login traces. Each API event is marked in gray; values in boxes following API events are their respective parameters and responses. Blue boxes mark \login passwords, yellow boxes mark the first \mfa codes, green boxes mark the second \mfa codes, and red boxes mark values that are the source of the property's violation. Traces annotated with a green check satisfy the property, while crossed traces are invalid. Edges annotated with green checks~\okmark{} and red crosses~\nomark{} denote whether the corresponding equality or inequality constraint is satisfied.}
\label{fig:mfa}
\end{figure}

For example, the first HTTP trace $(i)$ in \cref{fig:mfa} is valid. We write $\clogin$ and $\rlogin$ for call and return events of the \login API, and $\cmfa$ and $\rmfa$ for call and return events of \mfa API. The trace begins with a call to \login with password $1234$. In this execution, \login service initiates the first MFA challenge by calling \mfa API with challenge code $200$. This MFA challenge's response carries the same value, meaning that the challenge was successful. Then \login service initiates a second MFA challenge, but this time with a different code $100$. Once again, this MFA challenge also succeeds with the response carrying $100$. Thus, this trace satisfies the correct API event order and also the requirement that the two \mfa codes should be different from the password of the enclosing \login request, and that the call and return corresponding to a given \mfa request should carry the same code. 

Our policy does not require the two MFA challenge codes to be different from each other. For example, the trace $(ii)$ is also accepted. Meanwhile, trace $(iii)$ is rejected because the input code $200$ to the first \cmfa does not match its response $150$. Similarly, trace $(iv)$ is rejected because the first \mfa challenge reuses the \login password $1234$.

The SafeNom policy for the set of traces satisfying the above property is:

\[
\textcolor{safenomLogin}{
  \rotatebox[origin=c]{90}{\text{(c) login scope}}
}
\quad
\color{safenomLogin}\left\{
\color{black}
\begin{array}{@{}l@{}}
  \nu\namepilln{\bboxname{n}}.~
  \APIcall{login}~\namepilln{\bboxname{n}}
  \\[1.0ex]

  \quad
  \begin{array}{@{}l@{\quad}l@{}}
    \begin{array}{@{}l@{}}
      \left.
      \nu\namepillm{m}.~
      \APIcall{mfa}~\namepillm{m}
      \;\;
      \APIret{mfa}~\namepillm{\bboxname{m}}
      \color{safenomMFA}\right\}
      \;\textcolor{safenomMFA}{\text{(a)}}
      \\[0.75ex]

      \left.
      \nu\namepillmtwo{m}.~
      \APIcall{mfa}~\namepillmtwo{m}
      \;\;
      \APIret{mfa}~\namepillmtwo{\bboxname{m}}
      \color{safenomMFA}\right\}
      \;\textcolor{safenomMFA}{\text{(b)}}
    \end{array}
    &
    \textcolor{safenomMFA}{
    \begin{array}{@{}l@{}}
      \text{independent MFA scopes}\\
      \nu m\text{: fresh with respect to } \nu n
    \end{array}}
  \end{array}
  \\[1.0ex]

  \APIret{login}
\end{array}
\right.
\]

The correct sequence of HTTP requests/responses is marked in gray. The names $\namepilln{\bboxname{n}}, \namepillm{\bboxname{m}}$ are used  as logical identifiers for the API input and output parameters. The outer binder $\nu~\namepilln{\bboxname{n}}$ introduces a (fresh identifier) \textit{name} for the \login password. Its scope extends over the entire \login execution spanning from \clogin to its corresponding \rlogin. In a concrete trace, each binder is associated with the introduction of a fresh concrete value. Inside the \login scope, each MFA challenge introduces its own  local name $\namepillm{\bboxname{m}}$  for its respective code for the duration of its own execution. The policy has three crucial aspects: (a) the values of parameters marked by yellow $\namepillm{\bboxname{m}}$ within the first \mfa challenge should be equal, (b) the values of parameters marked by  green $\namepillmtwo{\bboxname{m}}$ within the second \mfa challenge should be equal, (c) (nested)  $\nu ~\namepilln{\bboxname{n}}$ and $\nu ~\namepillm{\bboxname{m}}$ implicitly specify that the values of parameters at positions marked with $\namepilln{\bboxname{n}}$ and $\namepillm{\bboxname{m}}$ should be distinct, but values of parameters in disjoint \mfa scopes need not be equal. Besides this, the policy should be agnostic to the actual value passed to all the parameters labeled by $\namepillm{\bboxname{m}}$ (and similarly by $\namepilln{\bboxname{n}}$).

\subsection{Traces as words with binders}
A concrete  microservice execution trace records the API events and the values that appear in their parameters and responses. For example, in trace (ii), the parameters/responses of both \mfa calls are $100$. But these values are indistinguishable and we cannot observe that these correspond to different binders. However, the policy needs to distinguish them. By inspecting the trace, one should be able to decipher that the first two $100$s belong to the first \mfa; and the last two belong to the second \mfa. Also, the trace does not record any information about the scopes in which the relevance of a value lasts for some (in)equality constraint. This is crucial for identifying that the first and second \mfa codes are not required to be equal.

To express these aspects, we add names $\namepilln{\bboxname{n}}$ and binders $\doublel \namepilln{\bboxname{n}}. \cdot \doubler$ to raw microservice traces. A name identifies which occurrences of concrete values in the trace are meant to be equal. A binder  $\doublel \namepilln{\bboxname{n}}. \cdot \doubler$ marks where the name $\namepilln{\bboxname{n}}$ is introduced and its associated concrete value's lifetime. Repeated occurrences of the name within its lifetime should have the same concrete value.   
For example, the MFA trace (ii) will be represented as:

\[
\left.
\begin{array}{@{}l@{}}
  \doublel\namepilln{n}.~\APIcall{login}~\namepilln{(\bboxname{n}, 1234)}
  \\[0.35ex]
  \quad
  \left.
  \begin{array}{@{}l@{}}
    \doublel\namepillm{m_1}.~
      \APIcall{mfa}~\namepillm{(\bboxname{m}_1, 100)}
      \;\;
      \APIret{mfa}~\namepillm{(\bboxname{m}_1, 100)}
    ~\doubler
    \\[0.35ex]
    \doublel\namepillmtwo{m_2}.~
      \APIcall{mfa}~\namepillmtwo{(\bboxname{m}_2, 200)}
      \;\;
      \APIret{mfa}~\namepillmtwo{(\bboxname{m}_2, 200)}
    ~\doubler
  \end{array}
  \right\}
  \ \begin{array}{@{}l@{}}
      \scriptsize\mfa\\[-0.45ex]
      \scriptsize\text{binders}
    \end{array}
  \\[0.35ex]
  \APIret{login}~\doubler
\end{array}
\right\}
\ \begin{array}{@{}l@{}}
    \scriptsize\text{login}\\[-0.45ex]
    \scriptsize\text{binder}
  \end{array}
\]

Here, $\namepilln{\bboxname{n}}$ is the name identifier for the \login password, and $\namepilln{\bboxname{m}_1}$ and $\namepillmtwo{\bboxname{m}_2}$ are the name identifiers for the first and the second \mfa codes. The two occurrences of $\namepillm{\bboxname{m}_1}$  must carry the same value $\namepillm{(\bboxname{m}_1, 100)}$. Similarly, the two occurrences of $\namepillmtwo{\bboxname{m}_2}$ should carry the same value $\namepillmtwo{(\bboxname{m}_2, 200)}$. Since $\namepillm{\bboxname{m}_1}$ and $\namepillmtwo{\bboxname{m}_2}$ are introduced by distinct binders with disjoint scopes, they can carry the same concrete values. However, $\namepillm{\bboxname{m}_1}$ and $\namepilln{\bboxname{n}}$ should carry distinct \mfa code and \login passwords as their binders are nested.

The binders and the names record the structure needed by the policy: (a) where a value is introduced, (b) where it should be reused or not, (c) when the constraints around it should stop being relevant. This trace model buys us a notion of renaming  concrete values while preserving the freshness and equality constraints on certain values. This is crucial for our policy's semantics which should only depend on equality and inequality repetition patterns instead of concrete values.
Therefore, \safenom uses nominal words with binders as its trace model. The \safenom policies are then interpreted over these nominal words with binders, rather than over flat sequences of concrete values.

\subsection{SafeNom policy enforcement}

A \safenom policy is enforced using a nominal automaton-based runtime monitor. Binders in traces offer a syntax-directed way to enforce the property. When the monitor sees a binder, it records the fresh value at that position under the bound name. When it later sees the same name again, it checks that the value matches the recorded one. When a new binder is introduced inside the scope of another name, the monitor checks that the new value is fresh with respect to the currently stored values. When the binder scope ends, the monitor can forget the value. Thus, the automaton does not need an ad-hoc implementation for each policy describing which values to remember, check, or drop.

The monitor simulating the nominal automaton, first, converts the raw microservice trace with flat API events and concrete values into a
nominal word by using symbolic finite
transducer \cite{sft}.\footnote{%
The \safenom policy to transducer compilation  is detailed in Appendix \appref{sec:transducer}.}
Then the nominal word output of the transducer is processed by the NA that
recognizes the given policy. If the NA accepts the nominal word, then the
corresponding raw microservice trace satisfies the property.

\section{Nominal Languages for Data-Aware Properties}\label{sec:extension}
To understand the interpretation of a data-aware property as a nominal language, we present a refresher on nominal languages followed by our nominal languages extension motivated by the microservice setting.
\subsection{Refresher: Nominal Languages with Binders}\label{subsec:intro-nom}
 
Nominal languages \cite{kurznomautomata} define sets of nominal words over a large (possibly infinite) domain of \textit{names} $\mathcal{N}$ and a finite set of \textit{letters} $\mathcal{C}$, where the names can be tested for equality. 
This is useful in applications, like microservices, where a possibly large set of parameter values, session tokens, etc., can be viewed as $\mathcal{N}$  and the set of function or API names can be viewed as $\mathcal{C}$.  Nominal words \cite{kurznomautomata} are given by the following grammar, where $\bboxname{c} \in \mathcal{C}$ and $\bboxname{n} \in \mathcal{N}$: 
\begin{align}\label{eq:nomgrammar}
	\text{Nominal words} \quad
	w ~::= ~ \epsilon ~\mid~\bboxname{c} ~\mid~\bboxname{n}~\mid~w \cdot w ~\mid \lazybindword{\bboxname{n}}{w}
\end{align}
The first four cases describe the ordinary trace structure: an empty trace, an event from the finite alphabet, a data value from the large \textit{name} domain, and sequential composition. The binder case $\lazybindword{\bboxname{n}}{w}$ is the key structure provided by nominal words. It introduces a scope for the name $\bboxname{n}$ in the word $w$. The occurrences of $\bboxname{n}$ inside this scope (\textit{i.e.,} inside the word $w$) are treated as occurrences of the same logical value (or name). 

To illustrate a nominal word, suppose that a successful \textit{file write} requires a \textsf{write} call to occur between an \textsf{open} and a \textsf{close} call, where \textsf{open} and \textsf{close} should be passed the same \sessid. We instantiate the grammar with session IDs as names $\mathcal{N}=\mathbb{N}$ and session actions as a finite set of symbols $\mathcal{C} = \{\textsf{open}, \textsf{close}, \textsf{write}\}$. 
A flat trace without any binder can only record the concrete session events and values:
\[\textsf{open}~100 ~\textsf{write}~\textsf{close}~100.\]
This trace only describes that the concrete value $100$ appears after both \textsf{open} and \textsf{close}. However, it does not record that the value after the  \textsf{open} and \textsf{close} events are supposed to be occurrences of the same logical value of \sessid and that only the $100$s between \textsf{open} and \textsf{close} events are to be treated as \sessid. We can express this using nominal words with binders as the following trace:
\[\doublel 100.~ \textsf{open}~100 ~\textsf{write}~\textsf{close}~100 \doubler.\]

Here, the binder $\doublel ~100. \_\doubler$ introduces a bound name $100 \in \mathcal{N}$. Within this scope,  all occurrences of $100$ are treated as uses of the same logical value, \textit{i.e.,} \sessid. Meanwhile,
\[\doublel 100.~ \textsf{open}~100 ~\textsf{write}~\textsf{close}~200 \doubler\] is a nominal word of an invalid trace, where \textsf{open} and \textsf{close} are called with different session IDs. 

Thus, nominal words with binders offer a promising foundation to record the three structural aspects of data-aware traces that are required to check the properties in our microservice settings: order of API events, the data values along with their positions, and the scope over which certain values are treated as the same logical value. 

The scoping aspect is a key tool for choosing nominal words with binders for interpreting our target properties that simply care about repetition of (in)equality patterns instead of concrete values in a run.     

A binder does not care about the concrete name it introduces. It only records the equality pattern among the occurrences of values in its scope. Therefore, the bound name is like a placeholder and replacing it consistently by another name gives an \textit{equivalent} trace. For instance, 
\[\doublel 200.~ \textsf{open}~200 ~\textsf{write}~\textsf{close}~200 \doubler\]
represents the same pattern as the above trace, but with a different \sessid $200$. With respect to the above property, both are equivalent valid traces because the values used at \textsf{close} is the same as that introduced at \textsf{open}. This consistent renaming of bound names in a nominal word is called $\alpha$-renaming. However, not every replacement of a bound name is valid.

\newpar{$\alpha$-renaming of nominal words}
To precisely state $\alpha$-renaming, we first distinguish \textit{bound} and \textit{free} occurrences of names in a nominal word. An occurrence of a name is \textit{bound} if it lies inside the scope of a binder that introduced the given name; otherwise it is \textit{free}. For example, in $w_1 =\doublel \bboxname{n}.~\bboxname{n}~\bboxname{n}~\bboxname{m}\doubler$, the two occurrences of $\bboxname{n}$ are bound by the initial binder, while $\bboxname{m}$ is \textit{free} as no binder binds it. Similarly, in the nominal word $w_2 = \doublel \bboxname{n}.~\bboxname{n}~\bboxname{n}\doubler~\bboxname{n}$, the two occurrences of $\bboxname{n}$ inside the binder delimiters are bound, but the final $\bboxname{n}$ is free as it is outside the binder's scope. 

$\alpha$-renaming changes the name introduced by a binder together with its occurrences bound by that binder. For instance, in $w_1$, we may rename  the bound name $\bboxname{n}$ to $\bboxname{p}$ as:
\[\doublel \bboxname{n}.~\bboxname{n}~\bboxname{n}~\bboxname{m}\doubler
\quad\equiv_\alpha\quad
\doublel \bboxname{p}.~\bboxname{p}~\bboxname{p}~\bboxname{m}\doubler.\]
These traces are said to be $\alpha$-\textit{equivalent} as the equality structure is unchanged: the first two names are bound occurrences of the same name and $\bboxname{m}$ remains free. The replacement name must not capture a freely occurring name. Thus, renaming $\bboxname{n}$ to $\bboxname{m}$ in $w_1$ is invalid:
\[
\doublel \bboxname{n}.~\bboxname{n}~\bboxname{n}~\bboxname{m}\doubler
\not\equiv_\alpha
\doublel \bboxname{m}.~\bboxname{m}~\bboxname{m}~\bboxname{m}\doubler .
\]
Here, the last occurrence of $\bboxname{m}$ was free in the original word, but it becomes bound in the renamed word. This changed the trace structure by adding an equality that was initially not present. 

Renaming does not affect occurrences of names outside the binder's scope. For instance, if we rename $\bboxname{n}$ to $\bboxname{p}$ in $w_2$, the last $\bboxname{n}$ remains unchanged:
\[
\doublel \bboxname{n}.~\bboxname{n}~\bboxname{n}\doubler~\bboxname{n}
\quad\equiv_\alpha\quad
\doublel \bboxname{p}.~\bboxname{p}~\bboxname{p}\doubler~\bboxname{n}.
\]

\newpar{Nominal Regular Expressions.}
We can express the set of valid nominal words using a nominal regular expression (NRE) \cite{kurznomautomata} given by the following grammar, where $\bboxname{c} \in \mathcal{C}$ and $\bboxname{n} \in \mathcal{N}$: 
\begin{align}\label{eq:orignre}
	ne ~::= 1~\mid~0~\mid~\bboxvar{c}~\mid~\bboxvar{n}~\mid~ne+ne~\mid ne \cdot ne~\mid ~ne^*~\mid~\bindexp{\bboxvar{n}}{ne} 
\end{align}
Here, $1$ matches the empty word $\epsilon$, $0$ matches no word, $\bboxvar{c}$ matches the same letter, and $\bboxvar{n}$ matches the same name. NREs also support the usual union, concatenation, and Kleene star. 

A \textit{binder} expression $\bindexp{\bboxvar{n}}{ne}$ matches words of the form $\doublel \bboxvar{n}. w \doubler$, where $w$ is matched by $ne$. This binder expression also matches the $\alpha$-renamings of $\doublel \bboxvar{n}. w \doubler$.

\begin{example}
Consider using the binder expression $\nu ~100. \textsf{open}~100
~\textsf{write}~\textsf{close}~100$ to specify the file write property. This
NRE matches $\doublel 100.~ \textsf{open}~100 ~\textsf{write}~\textsf{close}~100
\doubler$. Although the NRE includes $100$, it does not require that all the
matching nominal words should have \sessid set to $100$. The \sessid $100$ in
the above word could have been renamed to other names in $\mathcal{N}$. The key
point is that \emph{the parameter values following \textsf{open} and
\textsf{close} should be equal}. For instance, $\doublel 200.~ \textsf{open}~200 ~\textsf{write}~\textsf{close}~200 \doubler$ is also accepted by the policy. We could have written this policy using the NRE $\nu ~200. \textsf{open}~200
~\textsf{write}~\textsf{close}~200$ and yet it would have accepted the same set of valid traces. This invariance of the nominal language of an NRE under renaming of bound names is formally stated as the closure of nominal languages under $\alpha$-renaming.  
\end{example}

\newpar{Our extension.}
The nominal word accepted by the above NRE specification modeled the setting where the parameter value of \textsf{open} and \textsf{close} is observable at the binder, like ``$\doublel 100. $'', before the actual call. However, in reality, a function or API call's parameters are visible only after seeing the operation or API's identifier. The nominal words model by \kurz cannot express variants such as $\doublel \_.~ \textsf{open}~100 ~\textsf{write}~\textsf{close}~100 \doubler$, where the \sessid $100$ is unknown at the binder. This motivates our new notion of a lazy binder for modeling a microservice application's execution trace as a nominal word.

\subsection{Our Lazy Binding Extension}\label{subsec:lazy-nom-words}
To support lazy binding, we extend \kurz 's alphabet (comprising $\mathcal{N}$
and $\mathcal{C}$) with a set $ \uninitvar{\mathcal{N}}$ of \textit{partial}
names and a \textit{partial name} map $\eta: \mathcal{N} \to
\uninitnames{\mathcal{N}}$. Intuitively, a partial name is a logical identifier
for a parameter, while a name corresponds to the actual value of a parameter. Below, we denote a partial name as $\uninitvar{\bboxname{n}}$. We make two changes to the nominal words grammar in \cref{eq:nomgrammar}: (a) add partial names $\uninitvar{\bboxvar{n}} \in \uninitvar{\mathcal{N}}$, and (b) use partial names in binders, like $\doublel \uninitvar{\bboxname{n}}. \_\doubler$ instead of a name $\bboxname{n}$. 
\begin{align}\label{eq:lazygrammar}
	\text{Lazy nominal words} \quad
	w ~::= ~ \epsilon ~\mid~\bboxname{c} ~\mid~\bboxname{n}~\mid~\uninitvar{\bboxvar{n}}~\mid~w \cdot w ~\mid \lazybindword{\uninitvar{\bboxname{n}}}{w}
\end{align}
A name in the word between the binder delimiters is said to be \textit{bound} and \textit{corresponding} to $\uninitvar{\bboxname{n}}$ if the name is the first occurrence of a name that is mapped to $\uninitvar{\bboxname{n}}$ by $\eta$. For example, if $\eta(\bboxname{n}_1) = \uninitvar{\bboxname{n}}$, then in $w = \doublel \uninitvar{\bboxname{n}}. \bboxname{n}_1 \doubler \bboxname{n}_2$, 
the name $\bboxname{n}_1$ is the bound name corresponding to $\uninitvar{\bboxname{n}}$.

We note that the nominal word representation of real-world execution traces (in \S\ref{sec:casestudies}) will not have \textit{free partial names}, \textit{i.e.,} partial names appearing standalone without a binder $\doublel$ symbol. But our extended nominal words grammar allows free partial names as a technical device to define the semantics of our lazy NREs.

We also make changes to the standard NRE grammar in \cref{eq:orignre} and replace occurrences of names $\bboxvar{n}\in \mathcal{N}$ with partial names $\uninitvar{\bboxvar{n}}\in \uninitvar{\mathcal{N}}$: 
\begin{align}\label{eq:lazynre}
	\text{Lazy NRE}~ ne ~::= 1~\mid~0~\mid~\bboxvar{c}~\mid~\uninitvar{\bboxvar{n}}~\mid~ne+ne~\mid ne \cdot ne~\mid ~ne^*~\mid~\bindexp{\uninitvar{\bboxvar{n}}}{ne} 
\end{align}

The (lazy) binder expression in our grammar, $\bindexp{\uninitvar{\bboxvar{n}}}{ne}$, specifies a partial name $\uninitvar{\bboxvar{n}}$ and occurrences of $\uninitvar{\bboxvar{n}}$ in $ne$ mark the positions in a word $w$ matched by $ne$, where the name corresponding to $\uninitvar{\bboxvar{n}}$ should appear. 
The formal semantics of our lazy NREs is as follows:
\begin{definition}\label{def:lazy-nre-semantics}
 The set of nominal words accepted by a lazy NRE $ne$ is given by:
\begin{align*}
	 & L(1) \triangleq \{\epsilon\}, \quad                                                                                                                                                                                               
 L(0) \triangleq \emptyset                                                                                                                                                                                                                         , \quad 
L(\bboxvar{c}) \triangleq \{\bboxname{c}\},                                                                                                                                                                                                        \quad L(\uninitvar{\bboxvar{n}}) \triangleq \{\uninitvar{\bboxvar{n}}\}, \quad                                                                                                                                                                            
	 L(ne_{1} + ne_{2}) \triangleq L(ne_1) \cup L(ne_2)                                                                                                                                                                                                 \\
	 & L(ne_1 \cdot ne_2) \triangleq L(ne_1) \cdot L(ne_2) = \{w \cdot v ~\vert~ w \in L(ne_1), ~v \in L(ne_2)\}                                                                                                                                          \\
	 & L(ne^*) = \bigcup_{j \in \mathbb{N}} L(ne)^j, \text{where}~ L(ne)^{i+1} = L(ne) \cdot L(ne)^{i}, ~\text{and}~ L(ne)^0 = \{\epsilon\}\\
	 	 & L(\bindexp{\uninitvar{\bboxvar{n}}}{ne}) \triangleq 
\Set{ \lazybindword{\uninitvar{\bboxname{m}}}{w} \ | \begin{array}{l}
w' \in L(ne), ~\uninitvar{\bboxname{m}} \in \{\uninitvar{\bboxname{n}}\} \cup (\uninitvar{\mathcal{N}} \setminus FV(w')), \\ 
w = w'[\uninitvar{\bboxname{n}} \mapsto \bboxname{m}], ~\eta(\bboxname{m}) = \uninitvar{\bboxname{m}}
\end{array}}	 	 ~\text{where}
\end{align*}
 $FV(w')$ is the set of free partial names in $w'$ and 
the capture-avoiding substitution $w[\uninitvar{\bboxname{n}} \mapsto \bboxvar{m}]$ is defined as follows:
\[
w[\uninitvar{\bboxname{n}} \mapsto \bboxname{m}] =
\begin{cases}
w,
  & \text{if } w \in \{\epsilon,\bboxname{c},\bboxname{n}\} \\[0.4ex]

\bboxname{m},
  & \text{if } w = \uninitvar{\bboxname{n}} \\[0.4ex]

\uninitvar{\bboxname{k}},
  & \text{if } w = \uninitvar{\bboxname{k}} \text{ and }  \uninitvar{\bboxname{k}} \neq  \uninitvar{\bboxname{n}} \\[0.4ex]

w_1[\uninitvar{\bboxname{n}} \mapsto \bboxname{m}]
\cdot
w_2[\uninitvar{\bboxname{n}} \mapsto \bboxname{m}],
  & \text{if } w = w_1 \cdot w_2 \\[0.4ex]

\lazybindword{\uninitvar{\bboxname{n}}}{w'},
  & \text{if } w =
    \lazybindword{\uninitvar{\bboxname{n}}}{w'} \\[0.4ex]

\lazybindword{\uninitvar{\bboxname{k}}}
  {w'[\uninitvar{\bboxname{n}} \mapsto \bboxname{m}]},
  & \text{if } w =
    \lazybindword{\uninitvar{\bboxname{k}}}{w'}
    \text{ and } \uninitvar{\bboxname{k}} \neq \uninitvar{\bboxname{n}}.
\end{cases} \]
\end{definition}

Our semantics of the lazy binder expression differs from \kurz 's semantics. We allow the partial name $\uninitvar{\bboxname{n}}$ to be renamed to any partial name $\uninitvar{\bboxname{m}}$ not appearing freely, \textit{i.e.,} without a binder in a word $w' \in L(ne)$ matched by the inner $ne$. The word $w'$ will have free occurrences of $\uninitvar{\bboxname{n}}$ because it is not bound. Binding $\uninitvar{\bboxname{n}}$ substitutes all free $\uninitvar{\bboxname{n}}$ by a name $\bboxname{m}$ such that $\eta(\bboxname{m}) = \uninitvar{\bboxname{m}}$. 
\begin{example}
Consider the lazy NRE $ne = \nu~\uninitvar{\bboxname{n}}.  \uninitvar{\bboxname{n}} \uninitvar{\bboxname{n}}$. The binder marks that the first and second symbols in the word between $\doublel \uninitvar{\bboxname{n}}. \_\doubler$ should be some name $\bboxname{n}$ such that $\eta(\bboxname{n}) = \uninitvar{\bboxname{n}}$. For example, $ne$ matches $w = \doublel \uninitvar{\bboxname{n}}.\bboxname{n} \bboxname{n} \doubler$. Similar to the NREs (in \cref{eq:orignre}), our lazy NRE also accepts the renamed word $w' = \doublel \uninitvar{\bboxname{n}'}.~ \bboxname{n}' \bboxname{n}' \doubler$, where $\uninitvar{\bboxname{n}'} = \eta(\bboxname{n}')$. 
\end{example}
\begin{example} Consider $ne = (\nu~\uninitvar{\bboxname{n}}. \uninitvar{\bboxname{n}})^*$ and names $\bboxname{n}$ and $\bboxname{m}$, where $\eta(\bboxname{n}) = \uninitvar{\bboxname{n}} $ and $\eta(\bboxname{m}) = \uninitvar{\bboxname{m}}$. This NRE matches words with repetition, like $\doublel \uninitvar{\bboxname{n}}. \bboxname{n} \doubler$, $\doublel \uninitvar{\bboxname{n}}. \bboxname{n} \doubler \doublel \uninitvar{\bboxname{m}}. \bboxname{m} \doubler$, $\doublel \uninitvar{\bboxname{m}}. \bboxname{m} \doubler \doublel \uninitvar{\bboxname{m}}. \bboxname{m} \doubler$.
\end{example} 
\begin{example}
Consider   $ne = \nu \uninitvar{\bboxname{n}}. \nu \uninitvar{\bboxname{m}}. \uninitvar{\bboxname{m}} \uninitvar{\bboxname{n}}$ and names $\bboxname{n}$ and $\bboxname{m}$, where $\eta(\bboxname{n}) = \uninitvar{\bboxname{n}} $ and $\eta(\bboxname{m}) = \uninitvar{\bboxname{m}}$. This NRE matches $\doublel \uninitvar{\bboxname{n}}. \doublel \uninitvar{\bboxname{m}}. \bboxname{m} \bboxname{n}\doubler\doubler $. Permuting $\uninitvar{\bboxname{n}}$ and $\uninitvar{\bboxname{m}}$ gives another acceptable word $\doublel \uninitvar{\bboxname{m}}. \doublel \uninitvar{\bboxname{n}}.  \bboxname{n} \bboxname{m}\doubler\doubler$. 
\end{example}

Notice that NREs accept a language of \textit{closed} nominal words where every name  $\bboxname{n}$ in a word has a corresponding bound partial name $\eta(\bboxname{n})$. In the rest of the paper, by nominal words, we mean closed nominal words. 

\section{\textsf{SafeNom}: Nominal Language-based Policy Language}\label{sec:safenom}
Now, we turn to instantiating the general constructs of lazy NRE and nominal words
to the microservice setting.

We let $\mathcal{N} = \safenommodel$, so names are
pairs of a \emph{register} $n \in \regnametype$ and a \emph{value} $u \in \regvaltype$.
Intuitively, $u$ is the concrete value of an API parameter or output, while
the register $n$ is a \emph{logical identifier} describing the binder in the
policy that $u$ corresponds to. We let
the set of partial names be $\uninitvar{\mathcal{N}} =\{(n, \bot) \mid n \in
\mathcal{S}\}$ and the partial name map be $\eta((n, u)) = (n, \bot)$.
The set of symbols $\mathcal{C}$ contains API names prefixed with \textsf{call}
and \textsf{ret}, to denote API HTTP requests and responses.

However, we need one refinement to support inequality constraints in \safenom.
According to the lazy NRE semantics (in \cref{def:lazy-nre-semantics}), a nested binder policy, like $\nu (n, \bot). \nu (m, \bot). (n, \bot) (m, \bot)$ will accept both $\doublel (n, \bot). \doublel (m, \bot). (n, 100) (m, 200)\doubler \doubler$ and $\doublel (n, \bot). \doublel (m, \bot). (n, 100) (m, 100)\doubler \doubler$.  However, for a \safenom policy,  we want to enforce inequality between the value part of the names corresponding to the (nested) bound partial names, meaning $\doublel (n, \bot). \doublel (m, \bot). (n, 100) (m, 100)\doubler \doubler$ should be rejected under the \safenom semantics.

We show that we can capture inequality relations by restricting the semantics
of \safenom to a class of \emph{well-formed} words. As a further benefit, our
\safenom semantics satisfies a notion of $\alpha$-equivalence. 

\subsection{\safenom Semantics}\label{subsec:snom-semantics}
Intuitively, a well-formed word for \safenom should satisfy two constraints.
Firstly, each bound partial name must be associated with only one (register,
value) pair in the word. Secondly, to capture inequality, nested partial names
must be associated with  \emph{distinct} values. 

To formally define these constraints, we introduce a few notations. Consider a nominal word with nested binders, like $\doublel (n, \bot). (n, 1) \doublel (m, \bot). (m, 2) \doubler \doublel (r, \bot). (r, 2) \doubler \doubler$.

We call the set of partial names bound by nested binders between the outermost
and some innermost binder as a \textit{binder path}, and we let $Paths(w)$ be
the set of all paths in a word $w$.
For instance, in the above word the set of all paths is $\{\{(n, \bot), (m, \bot)\}, \{(n, \bot), (r, \bot)\}\}$.
For simplicity of presentation, we assume the bound partial names in \safenom
policies are all distinct; any \safenom policy can be rewritten to bring it to
this form.
We can now define well-formed words:

\begin{definition}\label{def:wellformedwords}
A nominal word $w$ is \emph{well-formed} if, for every path $p \in Paths(w)$ the following conditions hold.
\begin{enumerate}
\item \textit{\textbf{a unique name for any bound partial name:}} for any $(n_1, u_1), (n_2, u_2) \in w$, if $\eta((n_2, u_2)) = \eta((n_1, u_1)) =  (n_1, \bot) $ then $n_1 = n_2$ and $u_1 = u_2$.
\item \textit{\textbf{fresh nested names:}} the value parts (or the second projection) of the names corresponding to the bound partial names along the path $p$  are distinct. 
\end{enumerate}
\end{definition}

\begin{example}
The word $\doublel (n, \bot). (n, 1) \doublel (m, \bot). (m, 2) \doubler \doublel (r, \bot). (r, 2) \doubler \doubler$ is well-formed  as all  names of the form $(n, \_)$ have value $1$ and the value part of names $(n, 1), (m, 2)$ bound by the partial names along the path $\{(n, \bot), (m, \bot)\}$ are distinct. Similarly, along the other path, the names have distinct values $1$ and $2$. 
\end{example}

Finally, we are ready to define the \safenom semantics:
\begin{definition}
The set of nominal words accepted by a \safenom policy $ne$ is defined as $\snlang{ne} = L(ne) \cap \freshwords$, where $\freshwords$  is the set of well-formed nominal words.
\end{definition}

\subsection{Alpha-renaming in \safenom ~}\label{subsec:snom-alpharenaming}
Consider a \safenom policy that checks for equality between the first parameter
of API calls to \textsf{A} and \textsf{B}: $\nu (n, \bot). \textsf{callA}~ (n,
\bot) ~\textsf{retA~}\textsf{callB} ~(n, \bot) ~\textsf{retB}$. The acceptance
of a word should only depend on the equality of parameter values at certain
positions, rather than the specific values. We show this property holds for \safenom by proving that \safenom languages (like other nominal languages) are closed under a suitable notion of \alpharenaming.
 
We define \alpharenaming of a well-formed nominal word as the application of a
permutation (a bijection on $\mathcal{N}$), as usual:

\begin{definition}[Permutation Operation]
Applying a (bijective) permutation $\pi:\mathcal{N} \to \mathcal{N}$ to  a lazy nominal word $w$ is defined as:
\begin{align*}
& \pi(\epsilon) \triangleq \epsilon, \quad 
\pi(\bboxvar{c}) \triangleq \bboxvar{c}, 
\quad \pi(\uninitvar{\bboxvar{n}}) \triangleq \uninitvar{{\bboxvar{n}}}, \quad 
\pi(\bboxvar{n}) \triangleq \bboxvar{m}, ~\text{where} ~\bboxvar{n} \mapsto \bboxvar{m} \in \pi\\
& \pi(w_1 \cdot w_2) \triangleq \pi(w_1) \cdot \pi(w_2) \\
& \pi(\lazybindword{\uninitvar{\bboxname{n}}}{w}) \triangleq \begin{cases}\lazybindword{\uninitvar{\bboxname{m}}}{\pi(w)}, ~\text{if}~ \exists \bboxvar{n} \in w ~\text{s.t.}~ \eta(\bboxvar{n}) = \uninitvar{\bboxname{n}} ~\text{and}~\eta(\pi(\bboxvar{n})) = \uninitvar{\bboxname{m}} \\
\lazybindword{\uninitvar{\bboxname{n}}}{\pi(w)}, ~\text{otherwise}
\end{cases}
\end{align*}
\end{definition}

\begin{example}
Consider the well-formed word $\doublel (n, \bot). \doublel(m, \bot). \textsf{call-A}~(n, 100) ~\textsf{call-B}~(m, 200)\doubler \doubler$ capturing that API \textsf{A} is called with parameter value $100$ and \textsf{B} is called with parameter value $200$. When permuted using $\pi$, where $\pi(n, 100) = (m, 200)$ and $\pi(m, 200) = (n, 100)$, we get  $\doublel (m, \bot). \doublel(n, \bot). \textsf{call-A} ~(m, 200)~\textsf{call-B}~ (n, 100)\doubler \doubler$. This denotes another execution trace where \textsf{A} was called with value $200$ and \textsf{B} was called with $100$.
\end{example}
In the above example, applying a permutation to a well-formed word produces
another well-formed word.  However, applying the permutation with $\pi(n, 100)
= (n, 200)$ above gives:
\[\doublel (m, \bot).  \doublel(n, \bot). \textsf{call-A}
~(m, 200)~\textsf{call-B}~ (n, 200)\doubler \doubler,\] which is not well-formed.

Thus, to establish $\alpha$-equivalence, we must work with a restricted set of
permutations. To define these permutations, we introduce some notation.  We let
$\regname: \safenommodel \to \regnametype$ and $\regval: \safenommodel \to
\regvaltype$ denote the first and second projections. We define the restricted
domain of a path $p$ (the set of names corresponding to the partial names along
the path) to be $D_p = \{(n, u) \in w \mid (n, \bot) \in p~\text{and}~\eta((n,
u)) = (n, \bot)\}$, and we let $\regval(D_p) \triangleq\{u \mid (n, u) \in
D_p\}$. 

\begin{definition}[Valid \safenom Permutation]
Consider a well-formed nominal word $w$. Let the restricted permutation for any
$p \in Paths(w)$ be $\pi \upharpoonright_{D_p}: D_p \to \pi(D_p)$, the
restricted value and name projection functions be $\regval
\upharpoonright_{D_p}: D_p \to \regval(D_p)$ and $\regname
\upharpoonright_{D_p}: D_p \to \regname(D_p)$, respectively. A permutation $\pi$
is \emph{valid for  $w$} if for any path $p \in Paths(w)$:
\begin{enumerate}
\item the \textit{value transposition} $V_p: \regval(D_p) \to
  \regval(\pi(D_p))$ is a bijection:
\[
V_p(u) = \regval\upharpoonright_{D_p}(\pi \upharpoonright_{D_p}(\regval\upharpoonright_{D_p}^{-1}(u))),\]
\item the \textit{register transposition} $R_p: \regname(D_p) \to \regname(\pi(D_p))$ is a bijection:
\[
R_p(u) = \regname\upharpoonright_{D_p}(\pi \upharpoonright_{D_p}(\regname\upharpoonright_{D_p}^{-1}(u))).\]
\end{enumerate}
%Note that the $\regval\upharpoonright_{D_p}$ function is a bijection in case of a well-formed word.
\end{definition}

Note that the set of valid permutations depends on the word $w$. However, the
sets of valid permutations are closely related: there is a bijection that maps
every valid permutation for a well-formed word to a unique permutation that is
valid for the $\alpha$-renamed permuted word.
\begin{theorem}
Let $w_1$ be a well-formed word with some valid permutation $\pi \in \allperms{w_1}$, which permutes $w_1$ to $w_2 = \pi(w_1)$. The function $Perm: S(w_1) \to S(w_2)$ defined as $Perm(\pi_1) = \pi_2$, where $\pi_2(x) = \pi_1(\pi^{-1}(x))$ is a bijection.
\end{theorem}
We defer the proofs of this section's theorems to Appendix \appref{sec:defsec4}.

Applying a valid permutation to a word gives an $\alpha$-equivalent word:
\begin{definition}
  Well-formed words $w_1$ and $w_2$  are said to be \emph{$\alpha$-equivalent},
  \textit{i.e.,} $w_1 \equiv_{\alpha} w_2$ if there exists a valid permutation
  $\pi$ for $w_1$ such that $w_2 = \pi(w_1)$.
\end{definition}
As expected, $\equiv_{\alpha}$ is an equivalence relation:
\begin{theorem}
The $\equiv_{\alpha}$ relation is symmetric, reflexive and transitive.
\end{theorem}

Finally, we can show that \safenom policies are closed under
$\alpha$-equivalence.
\begin{theorem}
Let $ne$ be a \safenom policy and $w_1, w_2$ be two well-formed nominal words such that $w_1 \equiv_{\alpha} w_2$. If $w_1 \in \mathcal{L}(ne)$ then $w_2 \in \mathcal{L}(ne)$.
\end{theorem}
Thus,  \safenom policies do not depend on specific
parameter values of API calls and returns, but only on the (in)equality relation
across parameter values.

\section{Case Studies}\label{sec:casestudies}
In this section, we highlight the expressiveness of \safenom features using realistic case studies. 
Below, the API names are prefixed with \textsf{call} and \textsf{ret} to denote the API's request and response, respectively. 

We write $S$, where $S= \{s_1, \ldots, s_n\}$, as the shorthand for the regular
expression $(s_1 + \cdots + s_n)$ and $\excmark S$ for $(\mathcal{C} - S)^*$.

\newpar{Static data/configuration.}
\safenom can express properties where the exact values that appear at certain positions is statically known, like a Boolean error code, region identifiers, etc., using regular expression patterns over constants $\mathcal{C}$. 
For instance, consider a GDPR compliance regulation that mandates that the data of user in the EU region should not be stored in a US region's database to avoid violating  data storage guidelines. 
Consider the following two APIs in the database-backed application: API \APIname{User}, which takes the region of the user as its parameter, and API \APIname{DB} that takes in the region of the database to which it writes. 

The GDPR policy can be stated as: after a call to $\callUser$ with $\EU \in \mathcal{C}$, any call to \callDB should be passed $\EU$ to avoid cross-region writes. This is specified as follows:
\[\callUser~ \EU~(\excmark\{\callDB\} + \callDB ~ \EU)^*,\]
where $\excmark\{\callDB\}$ allows any other APIs to be invoked without any restrictions. This policy accepts the following trace:
\[\callUser~\EU ~ \callDB ~\EU~\retDB~\callDB~\EU~\retDB~\retUser,\]
but rejects the trace
\[\callUser~\EU ~ \callDB ~\US~\retDB~\callDB~ \retDB~\callDB~\retDB~\retUser\]
because the first $\APIname{DB}$ is writing to the $\US$ region and additionally the second $\APIname{DB}$ write did not carry the region identifier.

This class is useful when a policy depends on simple tags or flags. For example, a deployment team may use version tags for A/B testing to ensure that an end-to-end request is  served by the same API version; a security team may use traffic origin tags to prevent external requests from reaching confidential internal services; and a development team may route paid and unpaid users to different feature APIs based on some account-tier tags.
\newpar{Equality check.} 
Unlike the above policy over the constant region identifier, some properties require two or more API events to carry the same (runtime) parameters or responses. We express such properties using \safenom 's binder construct. Such policies arise when services must 
agree on dynamic values, like request identifiers, session tokens, transaction IDs, etc. For example, a tracing policy may require all downstream APIs to use the same trace ID (which will be assigned at runtime); an authorization policy may require all the database accesses to use the same session ID. 

As a representative example, consider the following APIs in a CI/CD build pipeline:
a build API $\APIname{Build}$ that takes in an identifier for the build session it starts; and a container creation API $\APIname{Container}$ that takes the identifier to tag the container's image. To avoid wastage of resources, a deployment team requires  that exactly one container image tagged with the same identifier as the build session should be created. In the following \safenom policy, $\hsvar{id}$ represents any identifier and all instances of $\hsvar{id}$  are required to have the same identifier: 
\[
\begin{array}{@{}l@{}}
  \nu~\hsvar{id}.~\APIcall{Build}~\hsvar{id} ~ 
    \\[0.35ex]
 \quad \begin{array}{@{}l@{}}
  	\excmark\{\APIcall{Container},~ \APIret{Container},~ \APIcall{Build},~ \APIret{Build}\}\\
    \APIcall{Container}~ \hsvar{id}~\APIret{Container}\\
    \excmark\{\APIcall{Container},~ \APIret{Container},~ \APIcall{Build},~ \APIret{Build}\}
  \end{array}
  \\[0.35ex]
  \APIret{Build}
\end{array}
\]

The pattern between \APIcall{Build} and \APIret{Build} disallows more than
one occurrence of $\APIcall{Container}$, and requires that occurrence to carry
the same \textsf{id} bound at \textsf{Build}. The following two traces
illustrate an accepted run (left) and a rejected run (right), where the
\textsf{Container} call either repeats or carries a mismatched value:
\[
\begin{array}{@{}c@{\qquad\qquad}c@{}}
\begin{array}{@{}l@{}}
  \doublel~\namepilldashedn{({id}, \bot)}.~\APIcall{Build}~\namepilln{({id}, 100)} ~ 
    \\[0.35ex]
 \quad \begin{array}{@{}l@{}}
    \APIcall{Container}~ \namepilln{({id}, 100)}~\APIret{Container}
  \end{array}
  \\[0.35ex]
  \APIret{Build} \doubler
\end{array}
&
\begin{array}{@{}l@{}}
  \doublel~\namepilldashedn{({id}, \bot)}.~\APIcall{Build}~\namepilln{({id}, 100)} ~ 
    \\[0.35ex]
 \quad \begin{array}{@{}l@{}}
    \APIcall{Container}~ \namepillbad{({id}, 200)}~\APIret{Container}\\
    \APIcall{Container}~ \namepilln{({id}, 100)}~\APIret{Container}
      \quad\text{\nomark}~{\scriptsize\text{extra}}
  \end{array}
  \\[0.35ex]
  \APIret{Build} \doubler
\end{array}
\\[0.8ex]
\text{\okmark}~\text{accepted} & \text{\nomark}~\text{rejected}\end{array}
\]

%\newpar{Binder in the middle:}
\newpar{Inequality check.}
Developers often want to check that an API generates fresh data, transforms an input before forwarding it, or returns  multiple distinct values. Such requirements appear when raw input to an encryption service should be different from its encrypted response, a backup copy should be stored in a different region than the primary region, or a user may create a new sub-task with a different session ID.  Such properties can be
stated as an inequality between nested bound names.
As a representative example, consider the following APIs that implement a secure handshake protocol: a frontend API \APIname{frontend} that initiates the handshake, a key generation API \APIname{KeyGen} that responds with a public key followed by a private key, an API \APIname{Client} that takes in a key as a parameter and publishes its input to the client, and a similar API \APIname{Server} that takes a key and publishes it to the server. 

Suppose the protocol requires that the public and private keys generated by
\APIname{KeyGen} should be unique and the client should publish the public key,
while the server should publish the private key. This can be specified as the
following policy, where the pattern of $\pubkey$ marks that the \retkeygen 's
public key  response should be equal to the key published to the client, and the
pattern of $\pvtkey$ marks that the \retkeygen 's private key response should be
equal to the key published to the server:
\[
\begin{array}{@{}l@{}}
  \nu~\pubkey. \\[0.35ex]
  \quad \nu~\pvtkey. \callfrontend \\[0.35ex]
 \qquad \begin{array}{@{}l@{}}
 	\callkeygen ~ \retkeygen ~ \pubkey ~ \pvtkey\\
 	\callpublish ~ \pubkey ~ \retpublish \\
 	 \calldecrypt ~ \pvtkey ~ \retdecrypt
  \end{array}
  \\[0.35ex]
  \retfrontend
\end{array}
\]
The nested binders implicitly  specify the inequality between $\pubkey$ ~ and $\pvtkey$. The following accepted trace on the left uses distinct public and private keys and each is published to the correct party. Meanwhile, the trace on the right is rejected because the public and the private key coincide.

\[
\resizebox{\linewidth}{!}{$
\begin{array}{@{}c@{\hspace{1em}}c@{}}
\begin{array}{@{}l@{}}
  \doublel~\namepilldashedn{(pub,\bot)}. \\[0.35ex]
  \quad \doublel~\namepilldashedm{(pvt,\bot)}.~\APIcall{frontend} \\[0.35ex]
  \qquad
  \begin{array}{@{}l@{}}
    \APIcall{KeyGen} ~ \APIret{KeyGen} ~
      \namepilln{(pub, A4)} ~ \namepillm{(pvt, C2)} \\
    \APIcall{Client} ~ \namepilln{(pub, A4)} ~ \APIret{Client} \\
    \APIcall{Server} ~ \namepillm{(pvt, C2)} ~ \APIret{Server}
  \end{array}
  \\[0.35ex]
  \APIret{frontend}~\doubler~\doubler
\end{array}
&
\begin{array}{@{}l@{}}
  \doublel~\namepilldashedn{(pub,\bot)}. \\[0.35ex]
  \quad \doublel~\namepilldashedm{(pvt,\bot)}.~\APIcall{frontend} \\[0.35ex]
  \qquad
  \begin{array}{@{}l@{}}
    \APIcall{KeyGen} ~ \APIret{KeyGen} ~
      \namepilln{(pub, A4)} ~ \namepillbad{(pvt, A4)} \\
    \APIcall{Client} ~ \namepilln{(pub, A4)} ~ \APIret{Client} \\
    \APIcall{Server} ~ \namepillbad{(pvt, C2)} ~ \APIret{Server}
  \end{array}
  \\[0.35ex]
  \APIret{frontend}~\doubler~\doubler
\end{array}
\\[0.8ex]
\text{\okmark}~\text{accepted}
&
\text{\nomark}~\text{rejected}
\end{array}
$}
\]
\newpar{Repeated (local) equality check.}
Running multiple iterations of an operation, each time with (short-lived) independent data is common in  applications. For instance, login retries with refreshed keys, multiple file transfers with per-file checksums, CI/CD build retries with their own build IDs, etc. The key here is that the equality constraint is local to one iteration of the operation. All APIs in the same iteration should agree on that iteration's value, but different iterations may use different local values. In \safenom , policies on one iteration's data can be stated using binders. This can be extended to a sequence of independent (per-iteration) constraints using a Kleene star.

Consider an authenticator application implemented using: frontend API \APIname{Auth} that initiates a login; API \APIname{Gen} that runs a random number generator and responds with a new short-lived PIN; API \APIname{Pin} that returns a new PIN by calling \APIname{Gen} and returning the PIN in \APIname{Gen}'s response. The authenticator is allowed to retry login by refreshing PINs. The retries can be captured using a Kleene star around $\hsvar{pin}$'s binder:
%\begin{align*}
%\textsf{callA}~(\nu~\hsvar{pin}.\textsf{callP}~\textsf{callG}~\textsf{retG}~\hsvar{pin}~\textsf{retP}~\hsvar{pin})^*~\textsf{retA}.
%\end{align*}
\[
\begin{array}{@{}l@{}}
  \APIcall{Auth}\\[0.35ex]
  (\nu~\hsvar{pin}. \APIcall{Pin} \\[0.35ex]
 \quad \begin{array}{@{}l@{}}
 	\APIcall{Generate}~\APIret{Generate}~\hsvar{pin}\\
  \end{array}
  \\[0.35ex]
  \APIret{Pin}~\hsvar{pin})^*\\[0.35ex]
  \APIret{Auth}
\end{array}
\]
Here, $\hsvar{pin}$ is bound for the scope of one PIN generation request to API
\APIname{Pin}: it must satisfy an equality constraint within each iteration, but it
may differ across iterations. For instance, the accepted trace on the left uses PIN $100$ in the first iteration and $200$ in the second, and in each iteration the same PIN that was sampled is passed through the \APIname{Generate} API.  Meanwhile, the second trace is rejected because the \APIname{Generate} API returned a PIN that is different from the one sampled by the \APIname{Pin} API.  
\[
\begin{array}{@{}c@{\qquad\qquad}c@{}}
\begin{array}{@{}l@{}}
  \APIcall{Auth}\\[0.35ex]
\begin{array}{@{}l@{}}
  \doublel \namepilldashedn{(pin, \bot)}. \APIcall{Pin} \\[0.35ex]
 \quad 
 	\APIcall{Generate}~\APIret{Generate}~\namepilln{(pin, 100)}\\
  \APIret{Pin}~\namepilln{(pin, 100)} \doubler \\
    \doublel \namepilln{(pin, \bot)}. \APIcall{Pin} \\[0.35ex]
 \quad 
 	\APIcall{Generate}~\APIret{Generate}~\namepilln{(pin, 200)}\\
  \APIret{Pin}~\namepilln{(pin, 200)} \doubler
  \end{array}
\\[0.35ex]
  \APIret{Auth}
\end{array}
&
\begin{array}{@{}l@{}}
  \APIcall{Auth}\\[0.35ex]
\begin{array}{@{}l@{}}
  \doublel \namepilldashedn{(pin, \bot)}. \APIcall{Pin} \\[0.35ex]
 \quad 
 	\APIcall{Generate}~\APIret{Generate}~\namepilln{(pin, 100)}\\
  \APIret{Pin}~\namepilln{(pin, 100)} \doubler \\
    \doublel \namepilln{(pin, \bot)}. \APIcall{Pin} \\[0.35ex]
 \quad 
 	\APIcall{Generate}~\APIret{Generate}~\namepilln{(pin, 200)}\\
  \APIret{Pin}~\namepillbad{(pin, 100)} \doubler
  \end{array}
\\[0.35ex]
  \APIret{Auth}
\end{array}
\\[0.8ex]
\text{\okmark}~\text{accepted} & \text{\nomark}~\text{rejected}
\end{array}
\]
\newpar{Repeated (global) equality check.}
Developers might want to express that certain data is unmutated or that some API call's response is the  same across invocations. For example, all calls in a transaction should carry the same transaction ID. In these cases, the value is introduced once and then reused many times. The number of reuses is not fixed in advance. \safenom can specify such properties as unrestricted repetition of some equal values using a Kleene star around the partial names. 
Consider the frontend API $\textsf{Frontend}$ that is the entry service of the application and two arbitrary APIs \textsf{A} and \textsf{B} that both take in a request ID parameter to identify the initial request for which the API calls are being invoked. 

For end-to-end observability, deployment teams rely on frontend $\textsf{Frontend}$ to carry a request identifier, which is passed unchanged to all downstream APIs during the execution to mark the APIs that are part of the same execution trace. We can specify this by binding a single $\hsvar{reqID}$ and repeatedly using it across all requests invoked while serving a call to \textsf{Frontend}:

\[
\begin{array}{@{}l@{}}
\nu~\hsvar{reqID}. ~\callTrace \\[0.35ex]
 \quad \begin{array}{@{}l@{}}
 	(\APIcall{A}~\hsvar{reqID} + \APIret{A}
 	+ \APIcall{B}~\hsvar{reqID}
 	+ \APIret{B}
 	+ \_)^*
  \end{array}
  \\[0.35ex]
  \retTrace
\end{array}
\]
The $\hsvar{reqID}$ is in scope for the entire execution. Using the Kleene star, we express that any call or return symbol can occur while serving \textsf{F} provided that the call symbols carry the $\hsvar{reqID}$ as their first parameter. The trace on the left satisfies this property as it consistently uses the same request ID. However, the trace on the right is rejected because the third child API to \textsf{B} does not carry a  request ID and also the request ID in the fourth child is different from the global identifier.

\[
\begin{array}{@{}c@{\qquad\qquad}c@{}}
\begin{array}{@{}l@{}}
\doublel~\namepilldashedn{(reqID, \bot)}. ~\callTrace \\[0.35ex]
 \quad \begin{array}{@{}l@{}}
 	\APIcall{A}~\namepilln{(reqID, 100)} ~\APIret{A}\\
 	\APIcall{A}~\namepilln{(reqID, 100)} ~\APIret{A}\\
 	 	\APIcall{B}~\namepilln{(reqID, 100)} ~\APIret{B}\\
 	 	 	\APIcall{A}~\namepilln{(reqID, 100)} ~\APIret{A}
  \end{array}
  \\[0.35ex]
  \retTrace \doubler
\end{array}
&
\begin{array}{@{}l@{}}
\doublel~\namepilldashedn{(reqID, \bot)}. ~\callTrace \\[0.35ex]
 \quad \begin{array}{@{}l@{}}
 	\APIcall{A}~\namepilln{(reqID, 100)} ~\APIret{A}\\
 	\APIcall{A}~\namepilln{(reqID, 100)} ~\APIret{A}\\
 	 	\APIcall{B} ~\APIret{B}~\quad\text{\nomark}\\
 	 	 	\APIcall{A}~\namepillbad{(reqID, 200)} ~\APIret{A}
  \end{array}
  \\[0.35ex]
  \retTrace \doubler
\end{array}
\\[0.8ex]
\text{\okmark}~\text{accepted} & \text{\nomark}~\text{rejected}
\end{array}
\]
\section{Enforcement: Lazy Nominal Automaton}\label{sec:enforcement}
Nominal languages are recognized by nominal automata. We extend \kurz 's nominal
automata (NA) model~\cite{kurznomautomata}, which supported only names and equality checks between names, to additionally support partial names for lazy binding and inequality checks on names. In this section, 
we work with non-deterministic NAs with $\epsilon$-transitions. In the following definition of our extended  (\textit{lazy}) nominal automata, we use the following notation $\setreg{i} \triangleq \{1, \ldots, i\}$, where $i \in \mathbb{N}$. 

\begin{definition}[Non-deterministic lazy $\mathcal{N}_{fv}$-nominal automaton]
Let  $\mathcal{N}$ be the set of names, $\uninitnames{\mathcal{N}}$ be the set of partial names, $\mathcal{C}$ be the set of constants, and $\mathcal{N}_{fv} \subseteq \uninitnames{\mathcal{N}}$ be a finite set of partial names. A $\mathcal{N}_{fv}$-nominal automaton is  a tuple $\mathcal{H} = (Q, q_0, F, \nomdelta, \mathcal{N}_{fv})$, where:
\begin{enumerate}
\item $Q$ is the finite set of states equipped with a map $\qtoregindex{\cdot}: Q \to \mathbb{N}$,
\item $q_0 \in Q$ is the initial state and $\qtoregindex{q_0} = 0$,
\item $F \subseteq Q$ is the set of final states and for all $q \in F$, $\qtoregindex{q} = 0$.
\item $\nomdelta: Q \times (I^{\mathcal{N}_{fv}}\cup \{\epsilon\}) \to 2^Q$, where $I^{\mathcal{N}_{fv}}= \mathcal{C} \cup \mathcal{N}_{fv} \cup \{i \in \setreg{\qtoregindex{q}}\mid q \in Q\} \cup \{\doublel, \doubler \} $,  and for any $q \in Q$ and $q' \in \nomtrans{q}{\alpha}$ the following holds:
\begin{align*}
& \alpha = \doublel \implies \qtoregindex{q'} = \qtoregindex{q} + 1\\
& \alpha = \doubler \implies \qtoregindex{q'} = \qtoregindex{q} - 1\\
&\alpha \in  (I^{\mathcal{N}_{fv}} \setminus \{\doublel, \doubler\}) \cup \{\epsilon\} \implies \qtoregindex{q'} = \qtoregindex{q}
\end{align*}
\end{enumerate}
\end{definition}
The states in NA are stratified into levels, where  $q$ is at \textit{level} $\qtoregindex{q}$. A transition from $q$ on $\doublel$ should go to a state in the next level. Similarly, transitions on $\doubler$ drop down to a state in the previous level. Transitions on constants, partial names in $\mathcal{N}_{fv}$, (state) levels, and $\epsilon$ preserve the level of the source state. 

A lazy nominal automaton is deterministic if it has a single initial state and for any $q$ and $\alpha$, $\nomtrans{q}{\alpha}$ is a singleton set.

Operationally, an NA processes a nominal word as a sequence of tokens of the
form `$\doublel \uninitvar{\bboxname{n}}.$', `$\doubler$' or symbols in
$\mathcal{N}, \mathcal{C}$, or $\uninitvar{\mathcal{N}}$, reading the tokens from left to right.  
An NA's configuration is a pair $\langle q, \sigma \rangle$ of a state $q$ and a store $\sigma: \setreg{\qtoregindex{q}} \to (\mathcal{N} \cup \uninitvar{\mathcal{N}}) \times 2^{\mathcal{N}}$. The store tracks the names currently in scope. 
An entry $\sigma(i) = (a, \forbidden)$ records that register $i$ currently holds the (partial) name $a \in \mathcal{N} \cup \uninitvar{\mathcal{N}}$ for (in)equality  checks, and that 
$a$ may later not be resolved to any names in the forbidden set $\forbidden$. We write $\fst(\sigma(i))$ and  $\snd(\sigma(i))$ for the first name component and the second forbidden set component. 

The small-step semantics of our lazy NA extended with inequality checks and lazy binding is defined as the following transition relation on configurations. Below, we use  $\sigma\upharpoonright_{\setreg{i}}$ to denote the restriction of $\sigma$ to the domain $\setreg{i}$. We write  $Img_{\fst}(\sigma) \triangleq \{a \mid (a, s) \in Img(\sigma)\}$ and similarly $Img_{\snd}(\sigma)$ for the second projections of $Img(\sigma)$. We write $\fresh{\bboxname{n}}{S}$ when the name $\bboxname{n}$ is fresh for every name in set of (partial) names $S$. For ordinary nominal words, this is $\bboxname{n} \notin S$. For \safenom names of the form $(n, u)$, the freshness check $\fresh{(n, u)}{S}$ holds iff for every name $(m, v) \in S$, we have $u \neq v$.

\begin{definition}[Single step transition]
	Given $q, q' \in Q$ and two configurations $t = \langle q, \sigma \rangle$ and $t' = \langle q', \sigma' \rangle$, a nominal automaton moves from $t$ to $t'$ on some symbol $s$, written $t \xrightarrow{s} t'$, 
	if there exists a \textit{transition symbol} $\alpha \in I^{\mathcal{N}_{fv}} \cup \{\epsilon\}$ such that $q' \in \nomtrans{q}{\alpha}$ and
	\[
		\begin{cases}
		\text{if}~s = \doublel\uninitnames{\bboxname{n}}. ~\text{then}~\alpha = \doublel~\text{and}~\sigma' = \sigma [\debruijn{q'} \mapsto (\uninitnames{\bboxname{n}}, \emptyset)] \\
\begin{array}{l}
		\text{if}~s = \bboxname{n} \in \mathcal{N} ~\text{then}~ 
		  		\alpha \in \setreg{\qtoregindex{q}} ~\text{s.t.}~ \forall j > \alpha.
        ~\eta^+(\fst(\sigma(\alpha))) \neq \eta^+(\fst(\sigma(j)))
        ~\text{and} \\
        \phantom{	\text{if}~s = \bboxname{n} \in \mathcal{N} ~\text{then}}
		\begin{cases}
			\sigma' = \sigma, ~\text{if}~ \fst(\sigma(\alpha)) = \bboxname{n}  \\
			\sigma' = \sigma[\alpha \mapsto (\bboxname{n}, \emptyset)],~\text{if}~ \fst(\sigma(\alpha)) =  \eta(\bboxname{n}), \\
			\qquad \fresh{\bboxname{n}}{Img_\fst(\sigma)}~\text{and}\\
			\qquad \fresh{\bboxname{n}}{\snd(\sigma(\alpha))}
		\end{cases}
		\end{array}
  \\
                    \text{if}~s = \doubler~\text{then}~\alpha=\doubler ~\text{and}
\begin{cases}
\sigma' = \sigma_{\setreg{\debruijn{q'}}}, ~\text{if}                                                            ~\fst(\sigma(\debruijn{q})) \in \uninitvar{\mathcal{N}}\\
\sigma' = \sigma{\setreg{\debruijn{q'}}},~\text{s.t.}\\
\begin{array}{l}
            \forall~ i \in [\debruijn{q'}], ~\fst(\sigma(i)) \in \uninitvar{\mathcal{N}} \implies
            ~\snd(\sigma'(i)) = \snd(\sigma(i)) \cup \{\fst(\sigma(\debruijn{q}))\}
\end{array}       
\end{cases}  \\
\text{if}~s = \epsilon ~\text{then}~\alpha = \epsilon~\text{and}~ \sigma' = \sigma\\
			\text{if}~s \in \mathcal{C} \cup \mathcal{N}_{fv} \setminus Img_\fst(\sigma) ~\text{then}~ \alpha = s~\text{and}~ \sigma' = \sigma                            \\
		\end{cases}
	\]
    where $\eta^+(\uninitvar{\bboxname{n}}) = \uninitvar{\bboxname{n}}$ and $\eta^+(\bboxname{n}) = \eta(\bboxname{n})$ for any $\uninitvar{\bboxname{n}}$ and $\bboxname{n}$.
\end{definition}
On reading a symbol $s$ at a configuration, the automaton picks a valid
transition symbol $\alpha$ to get the next configuration state. The store update during a transition depends on the token read.

\begin{enumerate}
\item \textbf{\textit{A transition on the lazy binder ``$\doublel\uninitnames{\bboxname{n}}$.''}} extends the store with the level of the new state $q'$ mapped to the partial name \bboxname{n}. The index of a (partial) name in the store marks the recency of reading the binder associated with the name, with a greater index denoting a more recent binder.
\item \textbf{\textit{A transition on $\bboxname{n}$}} captures the most
  important detail of our extended NA semantics for (in)equality checks on
  names. The automaton first resolves the binder corresponding to
  $\bboxname{n}$---the partial name associated with the binder should be the
  same as $\eta(\bboxname{n})$. Operationally, the NA searches for a transition
  symbol $\alpha$ such that either the name stored at $\sigma(\alpha)$, \textit{i.e.} $\fst(\sigma(\alpha))$ is $\bboxname{n}$ or
  $\uninitvar{\bboxname{n}}$. Since there might be many such indices, the NA resolves $\bboxname{n}$ to the most recent (innermost) binder;
  thereby supporting shadowing. In case $\fst(\sigma(\alpha)) = \uninitvar{\bboxname{n}}$, the NA also needs to ensure that the current name $\bboxname{n}$ being read is distinct from the names in: (a) the set $Img_{\fst}(\sigma)$ of names in the current store, and (b) the forbidden set for this register, \textit{i.e.,} $\snd(\sigma(\alpha))$. The freshness check against $Img_{\fst}(\sigma)$ prevents the new name from coinciding with other names that are currently in scope and, thus, in the registers.  The second check prevents the new name from coinciding with an old name that was bound in an inner scope that has ended. Such names no longer occur in the current store image and are therefore retained in the register's forbidden set. Once both  checks succeed, register $\alpha$ is updated to $(\bboxname{n}, \emptyset)$. The forbidden set is then discarded because once the name has been stored at a register, the role of the forbidden set is complete.
\item \textbf{\textit{A transition on $\doubler$}} closes the most recent binder scope, erases its register content, and updates the NA state. If the erased register contains a name, the NA adds that name to the forbidden set of every remaining register that holds a partial name. This propagation preserves the freshness constraint after the current bound name disappears from the current store. It is necessary because in the future, if a name is bound to one of the outer binders associated with the registers holding only partial names, the chosen name must be distinct from  all names bound in inner scopes during the outer binder's lifetime. Registers that already contain names are not updated because their contents are already fixed and will not undergo another freshness check.
\item \textbf{\textit{Transitions on constants, (free unbound) partial names and
  $\epsilon$}} only update the state. Note that the last transition rule handles (free) partial names in $\mathcal{N}_{fv}$ that are not bound (\textit{i.e.,} not in the current store's image).
 
\end{enumerate}

\begin{figure}
\centering
\begin{tikzpicture}[font=\scriptsize,
			baseline=1ex,shorten >=.4pt,node distance=25mm,on grid,
			semithick,auto,
			every state/.style={fill=white,draw=black,circular
					drop shadow,inner sep=.15mm,text=black,minimum size=1cm},
			accepting/.style={fill=gray,text=white}]
		\node (q0) [state, initial] {$q_0$};
		\node (q1) [state, right = of q0] {$q_1$};
        \node (q2) [state, right = of q1] {$q_2$};
		\node (q3) [state, right = of q2] {$q_3$};
        \node (q4) [state, right = of q3, accepting] {$q_4$};
		\path [-stealth, thick]
		(q0) edge [] node {\doublel}(q1)
		(q1) edge [] node {1}(q2)
		(q2) edge [] node {1}(q3)
    (q3) edge [] node {\doubler}(q4)
		;
	\end{tikzpicture}
	\[
    \langle q_0, \{\} \rangle \xrightarrow[s=\doublel (n, \bot)]{\alpha=\doublel}  \langle q_1, \{1 \mapsto ((n, \bot), \emptyset)\} \rangle \xrightarrow[s=(n, 100)]{\alpha=1} \langle q_2, \{1 \mapsto ((n, 100), \emptyset)\} \rangle \xrightarrow{} \ldots \xrightarrow{} \langle q_4, \{\} \rangle
  \]
 \caption{Nominal automaton that accepts $\snlang{\nu~(n, \bot). (n,100) (n,100)}$.}
 \label{fig:eg-na}
\end{figure}
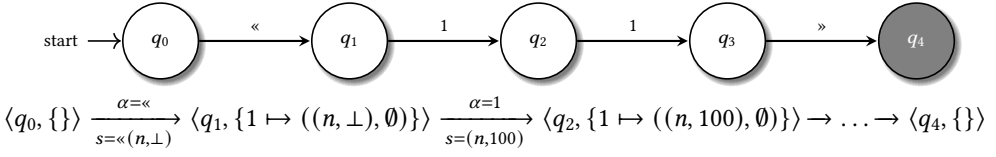

\begin{example}

Consider the NA $\mathcal{H}_{ne}$ in \cref{fig:eg-na} for the following \safenom ~ policy: $ne = \nu~(n, \bot). (n,1) (n,1)$. Here, $q_0$ is the initial state and $q_4$ is the final state, and a transition labeled with $\alpha$ from some state $q_i$ to another $q_j$ denotes $q_j \in \nomtrans{q_i}{\alpha}$. In the run of the word $\doublel (n, \bot). (n, 100) (n, 100)\doubler$, the NA first transitions from the initial state on the symbol $\alpha = \doublel$ and adds the partial name $(n, \bot)$ and empty forbidden set in the store indexed by the binder's level, \textit{i.e.,} $1$; then on the name $(n, 100)$, the NA transitions on symbol $\alpha=1$, which is the store's index mapping to the partial name corresponding to $(n, 100)$; and so on. Finally, the NA accepts the word by arriving at the final state with an empty store.

\end{example}

The following example illustrates how the forbidden sets enforce freshness with respect to names whose scopes have already ended.
\begin{example}
Consider the policy:
$
\nu~(n,\bot).(\nu~(m,\bot).~(m,\bot))~(n,\bot),
$
which requires the names bound to $(n,\bot)$ and $(m,\bot)$ to be distinct. Importantly, in a trace, a name will be bound to the outer binder only after the scope of the inner binder has ended. 
Consider the word:
\[
\doublel~(n,\bot).\doublel~(m,\bot).~(m,100)~\doubler(n,200)\doubler.\]
 Suppose after reading the two binders, the store contains $
  \sigma(1)=((n,\bot),\emptyset)$ and
$
  \sigma(2)=((m,\bot),\emptyset)$.  Reading $(m,100)$ will update the second register to $\sigma(2)=((m,100),\emptyset)$. When the inner scope closes, the NA will erase the second register. Since the first register still holds a partial name, $(m,100)$ will be added to its forbidden set, resulting in  $\sigma(1)=((n,\bot),\{(m, 100)\})$. The subsequent transition on $(n,200)$ will succeed because it is distinct from every name in the forbidden set. In contrast, the word $\doublel(n,\bot).\,
    \doublel(m,\bot).\, (m,100)\,\doubler\,
    (n,100)\,
  \doubler$ is rejected because the freshness check against the forbidden set is violated. However, if our freshness check was limited to only the currently stored names in the registers, we would have incorrectly accepted the trace.  This  example highlights a subtlety of supporting freshness with lazy binders.
\end{example}

We write $\mathcal{L}(\mathcal{H})$ for the language of a nominal automaton
$\mathcal{H}$: the set of all accepted words. These languages are closed under
$\alpha$-equivalence:
\begin{theorem}
Consider a lazy $\mathcal{N}_{fv}$-nominal automaton $\mathcal{H}$. For any two $\alpha$-equivalent well-formed words $w, w'$, if $w \in \mathcal{L}(\mathcal{H})$ then $w' \in \mathcal{L}(\mathcal{H})$.
\end{theorem}
The proof for this theorem and the others in this section are in the Appendix.

\newpar{Our \safenom policy to NA compiler.}
A \safenom policy is compiled into a deterministic NA using \kurz 's Thompson-style inductive construction of an NA from an NRE. Our compilation details can be found in Appendix \appref{sec:defsec6}; here, we summarize its main cases. Let $\mathcal{H}_{ne}$ denote the NA constructed for a policy $ne$.
\begin{itemize}
\item $\mathcal{H}_{0}$ is a single-state NA with no transitions and no final state. $\mathcal{H}_{1}$ is a single state NA with no transitions and a final state that is the same as the initial state.
\item Each of $\mathcal{H}_{\bboxname{c}}$ and $\mathcal{H}_{\uninitvar{\bboxname{n}}}$ is a  two-state NA with a singleton transition on $\bboxname{c}$ and $\uninitvar{\bboxname{n}}$,  respectively.
\item For the binder case $\mathcal{H}_{\nu \uninitvar{\bboxname{n}}. ne}$, the construction introduces new initial and final states. The new initial state enters $\mathcal{H}_{ne}$ on `` $\doublel$'' and the final state of $\mathcal{H}_{ne}$ transitions to the new final state on $\doubler$. Within $\mathcal{H}_{ne}$, every transition labeled by the bound partial
name $\uninitvar{\bboxname{n}}$ is replaced by a transition labeled $1$, the register level of the newly introduced binder. Wrapping $ne$ in this binder shifts every binder already occurring inside $ne$ one level deeper. Accordingly, each existing transition labeled by a level $i$ is relabeled
by $i+1$.
\item For the union, concatenation, and Kleene star cases, the construction goes through a Thompson-style construction \cite{thompson}.
\end{itemize}
Finally, the resulting non-deterministic NA is converted into a deterministic
NA using \kurz's layer-wise powerset construction.

As expected, our compilation procedure is sound:
\begin{theorem}\label{thm:soundness}
  Let $ne$ be a \safenom ~ policy. Then $\snlang{ne} = \mathcal{L}(\mathcal{H}_{ne})$, \textit{i.e,} $w$ matches $ne$ iff $\mathcal{H}_{ne}$ accepts $w$.
\end{theorem}

\section{Implementation}\label{sec:impl} 
To demonstrate our policy checking design for a microservice application, we
implemented a prototype in $\sim 3000$ lines of Java. Given a \safenom policy, our tool compiles a nominal automaton. Then the tool extracts runtime monitors, each of which simulates the compiled NA's transitions atop the Istio service mesh \cite{istio}, a new networking layer. Each service has a co-located monitor, as shown in \cref{fig:mondiag}, running outside the service without any invasive changes to the service  implementations. The runtime monitoring is \textit{distributed} across the (local monitors at) services. The monitor at a service runs when the given service sends or receives an HTTP message. Our monitor's Istio-specific design aspects are described below.

\newpar{Istio-based design.}
Service mesh design enables blackbox enforcement of our safety properties. In an application deployed on top of the Istio service mesh, each microservice runs in an independent (service) container and is paired with a co-located sidecar container that
implements an Envoy proxy \cite{envoydoc}. The proxy intercepts all HTTP
requests and responses corresponding to its service and can read the HTTP
message headers, query parameters, response values, add/delete/update the
headers, or allow/block HTTP messages. We implement our monitor in the proxy
using a WebAssembly Envoy filter \cite{envoyfilter}. These filters can be dynamically plugged into the sidecar proxy. Thus, \safenom can be deployed dynamically without modifying service implementations.

\newpar{From HTTP events to nominal words}
The proxy can only observe
API call/return events and the parameters, but the NA needs its input to be nominal
words. To bridge this gap, we insert a lightweight parsing stage in the monitor to reconstruct the implicit nominal structure in the trace before running the nominal automaton. The parser is a symbolic finite-state transducer compiled from the SafeNom policy being enforced.  The transducer's role is to simply parse the flat sequence of API events, extract the relevant parameter values, and insert nominal tokens, like binder delimiters at positions specified by the policy. 
 The transducer does not perform the semantic checks for equality, freshness, or scope; these checks are performed by the NA. Thus, the monitoring pipeline separates parsing from enforcement. 
 
Consider a policy requiring the ID received by \APIname{Frontend} to be the same as the ID used to look up data in a database:
\[
\begin{array}{@{}l@{}}
\nu~\svar{id}.~ \\[0.35ex]
\quad \APIcall{Front}~\svar{id}~~\\[0.35ex]
\quad \APIcall{DB}~\svar{id}~~
\APIret{DB}~~\\[0.35ex]
\APIret{Front}\ldots
\end{array}.
\]
Suppose $\APIcall{Front}$ received input $100$. Since the policy says that $\svar{id}$ gets bound at $\APIcall{Front}$, the transducer emits an opening binding delimiter $\doublel \namepilln{(id, \bot)}.$, followed by $\APIcall{Front}$; then the policy specifies that the name following $\APIcall{Front}$ should be bound to $\svar{id}$, so the transducer parses the value $100$ observed in the raw trace and  emits $\namepilln{(id, 100)}$. At each step the transducer updates its local state. These four tokens then get fed to the NA located at the sidecar proxy at the $\APIname{Front}$  service. The NA updates its states and pushes the name $\namepilln{(id, 100)}$ in the first register. This separation is useful as the transducer handles the syntactic parsing while the NA handles the policy checking. 

\newpar{Monitor configuration}
The monitor's configuration---transducer state, the NA state, and the NA store---is carried as custom HTTP headers as shown in \cref{fig:mondiag}. When an HTTP message arrives at a monitor, the transducer and the NA run in lockstep.  First, the transducer reads its current state from the HTTP headers and transitions on the sequence of API names and parameter values carried in the message. The value of the transducer's old state header is updated with the transducer's new state and the transducer's output tokens are fed to the NA. 
Then, the NA reads its current configuration headers and transitions on the input tokens. The value of the NA's old state header is updated with the NA's new state.

The update to the headers encoding the NA store  requires some care. On reading a token of the form ``$\doublel (n,\bot).$'', the monitor adds a new header named $n$ and sets its value to $\bot$. This corresponds to extending the store with $(n, \bot)$. Meanwhile, on reading $\doubler$, the most recently added header in the store is removed. On reading a parameter value, the relevant store headers are compared for (in)equality. 
Note that the services are responsible for propagating the headers from the parent to the child request/response for distributed monitoring.

\newpar{Deterministic transducer}
Since our monitor is online, the transducer must be deterministic because the transducer cannot backtrack to change its output. After reading a prefix of the trace, the transducer should determine what symbol to output simply by looking at the next event. For example, consider the above policy were part of some bigger policy, and below is a small sub-expression of that policy:
\[
\begin{array}{@{}l@{}}
\ldots~(\APIcall{Front}~\APIret{Front})^* \\[0.35ex]
\begin{array}{@{}l@{}}
\nu~\svar{id}.~ \\[0.35ex]
\quad \APIcall{Front}~\svar{id}~~\\[0.35ex]
\quad \APIcall{DB}~\svar{id}~~
\APIret{DB}~~\\[0.35ex]
\APIret{Frontend}\ldots
\end{array}
\end{array}
\]
Here, $\APIname{Front}$ may be invoked repeatedly, but starting from some intermediate invocation, the monitor must bind the name $\svar{id}$. However, a transducer  cannot determine the occurrence of $\APIcall{Front}$  at which it should introduce the binder for $\mathhighlight{\svar{id}}$. Disambiguating the position matched by a symbol without more than one-symbol lookahead is central for an online transducer to emit the binder delimiters at correct positions and correctly parse the required data from the API parameters.

To avoid such cases, we enforce a syntactic well-formedness requirement on policies used for online monitoring. 
The condition is stated on the sequence of observable API events obtained by erasing all binder annotations from the policy. Intuitively, the regular expression skeleton we get after ignoring the binder annotations, must have a unique left-to-right parse. Formally, the binder-free regular expression skeleton of the \safenom policy being monitored should be one-unambiguous. One-unambiguity  lets the transducer deterministically choose the position in the NRE from which the currently read symbol has been generated.  The benefit of this condition is that it makes the monitor frontend deterministic by construction while  keeping the NA's semantics unchanged. This is crucial for an online monitor. We can statically check the well-formedness of a property using the definition in Appendix. Also, a detailed explanation of the transducer  construction is given in the Appendix.

\newpar{Rejecting invalid traces.} For simplicity, our prototype logs any policy violation by checking if after processing the trace, the NA is in an accepting configuration or not. However, it is possible to actively block intermediate  HTTP messages as soon as the NA's rejecting  configuration is reached.  

\tikzset{
  meshservice/.style={
    draw=pwblue!70, thick, rounded corners=8pt,
    fill=pwblue!8, text=pwblue!80!black,
    minimum width=3.0cm, minimum height=1.2cm,
    align=center, font=\bfseries\small,
    drop shadow={shadow xshift=0.6pt, shadow yshift=-0.6pt, opacity=0.25}
  },
  meshmonitor/.style={
    draw=mfateal!80!black, thick, rounded corners=8pt,
    fill=white, text=mfateal!72!black,
    minimum width=3.0cm, minimum height=1.2cm,
    align=center, font=\small,
    drop shadow={shadow xshift=0.6pt, shadow yshift=-0.6pt, opacity=0.2}
  },
  meshannot/.style={
    draw=namecol!55, fill=namecol!7, rounded corners=3pt,
    font=\scriptsize, align=left,
    inner sep=4pt, text width=2.0cm, text=black!75
  },
  meshstep/.style={
    circle, draw=pwblue!70, thick, fill=white, text=pwblue!80!black,
    minimum size=0.54cm, font=\small\bfseries, inner sep=1pt
  },
  meshdocker/.style={
    fill=pwblue!75!black, rounded corners=2pt,
    font=\tiny\bfseries, text=white,
    inner sep=2pt, anchor=north east
  },
  meshenvoy/.style={
    fill=mfateal!75!black, rounded corners=2pt,
    font=\tiny\bfseries, text=white,
    inner sep=2pt, anchor=north east
  }
}
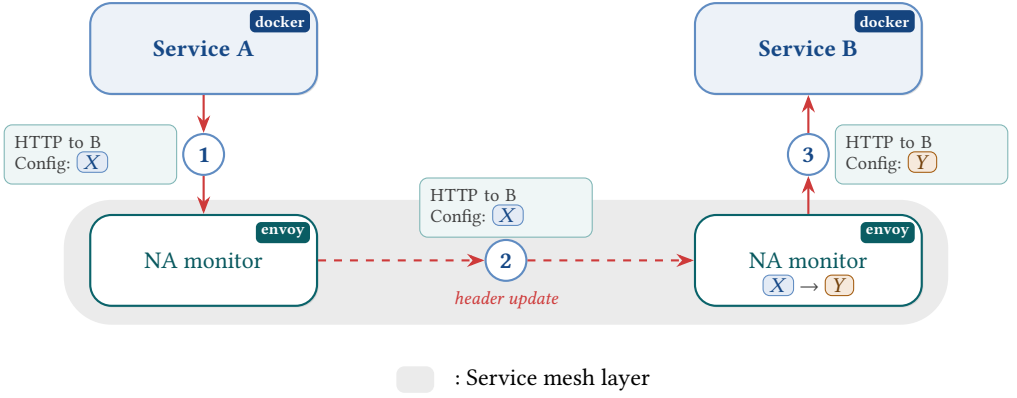
\begin{figure}
\begin{tikzpicture}[>=Stealth]

\node[meshservice] (serviceA) at (0, 0)    {Service A};
\node[meshservice] (serviceB) at (8, 0)    {Service B};
\node[meshmonitor] (monitorL) at (0, -2.8) {NA monitor};

\node[meshmonitor] (monitorR) at (8, -2.8) {NA monitor};
\node[font=\scriptsize, text=black!75] at ($(monitorR.center) + (0,-0.33)$)
    {$\namepilln{X}\to\namepillm{Y}$};

\node[meshdocker] at ($(serviceA.north east) + (-0.10, -0.10)$) {docker};
\node[meshdocker] at ($(serviceB.north east) + (-0.10, -0.10)$) {docker};

\node[meshenvoy] at ($(monitorL.north east) + (-0.10, -0.10)$) {envoy};
\node[meshenvoy] at ($(monitorR.north east) + (-0.10, -0.10)$) {envoy};

\begin{scope}[on background layer]
    \fill[black!8, rounded corners=18pt]
        (-1.85, -3.65) rectangle (9.85, -2.0);
\end{scope}

\node[meshstep] (n1) at (0, -1.4) {1};
\draw[badred, ->, thick] (serviceA.south) -- (n1.north);
\draw[badred, ->, thick] (n1.south)       -- (monitorL.north);
\node[meshannot, anchor=east] at (-0.35, -1.4)
    {HTTP to B\\Config: $\namepilln{X}$};

\node[meshstep] (n2) at (4.0, -2.8) {2};
\draw[badred, dashed, ->, thick] (monitorL.east) -- (n2.west);
\draw[badred, dashed, ->, thick] (n2.east)       -- (monitorR.west);
\node[meshannot, anchor=south] at (4.0, -2.5)
    {HTTP to B\\Config: $\namepilln{X}$};
\node[freshlbl, font=\scriptsize\itshape, text=badred, anchor=north] at (4.0, -3.08)
    {header update};

\node[meshstep] (n4) at (8, -1.4) {3};
\draw[badred, ->, thick] (monitorR.north) -- (n4.south);
\draw[badred, ->, thick] (n4.north)       -- (serviceB.south);
\node[meshannot, anchor=west] at (8.35, -1.4)
    {HTTP to B\\Config: $\namepillm{Y}$};

\fill[black!8, rounded corners=3pt]
    (2.55, -4.55) rectangle (3.05, -4.2);
\node[anchor=west, font=\small] at (3.2, -4.375)
    {: Service mesh layer};
\end{tikzpicture}
\caption{Blackbox deployment of \safenom monitor. Each service runs in its own Docker container and the  service mesh layer pairs each container with a sidecar container running Envoy proxy. When a request from \APIname{A} is sent to \APIname{B} in step 1,  the HTTP header carries the current monitor configuration. Upon arrival at  \APIname{B}'s sidecar proxy, the monitor updates the configuration in step 2, and forwards it to the service container in step 3.}
\label{fig:mondiag}
\end{figure}

\section{Evaluation}\label{sec:eval}

We evaluate \safenom monitor's memory footprint and performance overhead by answering the following research questions:
\begin{itemize}
\item \textit{RQ1: How much header space is required for the configuration headers?}
\item \textit{RQ2: How much latency overhead does the monitor add?}
\end{itemize}

We evaluate SafeNom on a suite of policies that cover the main language constructs. The \safenom monitor is deployed in an Istio-enabled Kubernetes cluster. We instantiate the policy suite on two Go-based microservice applications: a hospital workflow and a hotel reservation application \cite{deathstar} running in the cluster. The average number of nodes in the service tree of both applications is 6 and 4.5, respectively. Since \safenom runs outside the application, its performance overhead is not affected by the application's internal implementation. However, it depends on the policies being checked, which we evaluate. The case studies and the detailed deployment information are in Appendix \appref{sec:deployment}.

\subsection*{RQ1: Header Space Overhead}
Carrying the NA's and transducer's current configuration is key to our
enforcement mechanism. \Cref{tbl:eval} presents (for each policy in the Policy
column) the total number of NA states (\#nstates); number of bits for encoding
the maximum number of NA states (\#nbits) and transducer states (\#tbits);
number of NA transitions to non-rejecting state (\#ntrans); and the maximum number
of headers for the NA store (\#levels). All our policies, ranging from no
nesting to three levels of nesting, require at most twelve bits of header space for the NA and transducer state. In our experiments, we consider 32-bit parameters. So the number of bits for carrying the NA store is equal to the $\#levels \times 32$. This is minimal compared to the available space of HTTP headers (on the order of kilobytes).  
\textbf{
\textit{To summarize, the \safenom monitor compactly encodes its configuration into a few bits of extra  header space. }}

\begin{table}[t]
\caption{Policies prefixed with ``Hotel'' are evaluated on the hotel
  application, and the remainder on the hospital application. Main
  findings: (1) monitor configuration can be encoded in a few bits of extra header space, and (2) monitoring adds minimal latency, on the order of milliseconds to the application.}

\label{tbl:eval}
\centering
\normalsize
{\begin{tabular}{ l c ccccccc }
\toprule
  & \multicolumn{3}{c}{NA}     & Max. Nesting & Transducer &Latency\\
\emph{Policy}  & \#nstates & \#nbits & \#ntrans
            & \#levels & \#tbits & overhead (ms)\\
\midrule
 1.  EU Data & 5 & 3 & 17      &   0 & 5 & 1.08\\
 2.  CI/CD &    10  & 4 & 33   &  1  & 6 & 0.86\\
  3. Propagate Trace ID & 10 &4 &  21      &  1  & 5 & 1.26\\
  4.  Data Encryption& 15& 4& 26     & 2  & 5 & 0.50\\
 5. Key & 17 & 5 &   16     &2 & 5  & 0.329\\
 6. Session Auth & 10 &4& 10      &  2 & 4 & 0.83\\
 7. Three levels & 12 &4 &  11     &  3 & 4 & 0.55\\
 8. Hotel Data Encryption& 15& 4& 26     & 2  & 5 & 0.800\\
 9. Hotel EU Data & 5 & 3 & 17      &   0 & 5 & 0.45 \\
10. Hotel Three levels & 12 &4 &  11     &  3 & 4 & 0.61\\

    \bottomrule
\end{tabular}
}
\end{table}

\subsection*{RQ2: Latency Overhead}
To measure \safenom 's impact on the application's performance, we compare the latency of requests when the application is being monitored versus when it is not on a workload of 200 requests for all user-facing endpoints in our benchmark applications. We average the latency over five such workloads. Our experiment setup involves extracting Envoy filters from each policy's NA and then measuring latency overhead when the filter is enabled versus disabled. 
\newpar{Overhead versus policy.}
The average latency overhead  of monitoring each policy in  \cref{tbl:eval} is reported in milliseconds in \cref{tbl:eval}'s ``Latency overhead'' column. The policies prefixed with ``Hotel'' were evaluated on the hotel application and others were evaluated on the hospital application. Observe that across both applications, the values are consistently at most $\sim 1.5$ms. This is expected since the internal service implementation should not impact the monitoring overhead. \textit{\textbf{To conclude, the  \safenom monitor adds minimal latency overhead on the order of a millisecond. }}

\begin{figure}[t]
    \centering
    \begin{subfigure}[t]{0.48\textwidth}
        \centering
        \includegraphics[height=3.3cm,width=0.92\linewidth]{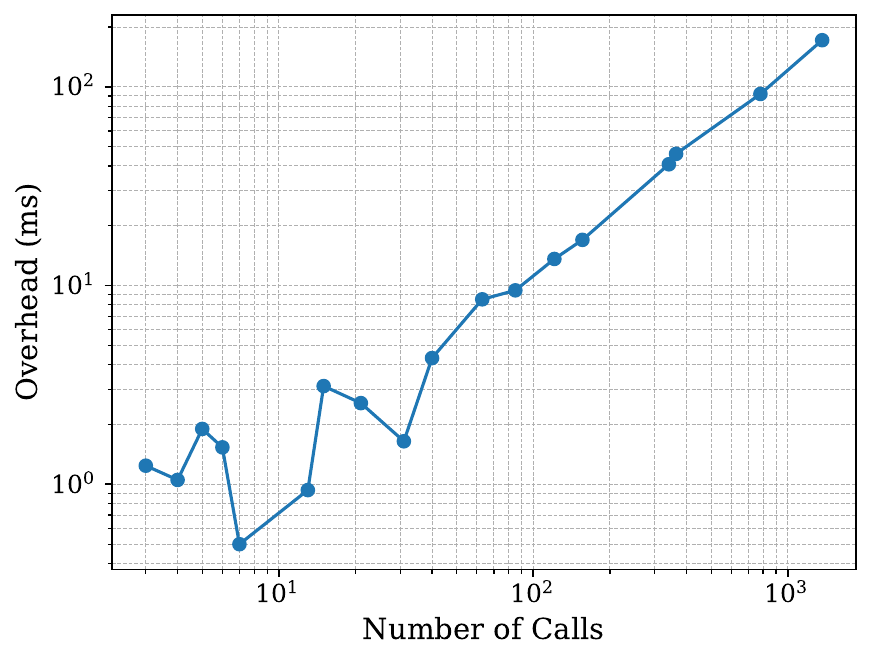}
        \caption{Overhead vs Calls}
        \label{fig:scaleall}
    \end{subfigure}
    \hfill
    \begin{subfigure}[t]{0.48\textwidth}
        \centering
        \includegraphics[height=3.3cm,width=0.92\linewidth]{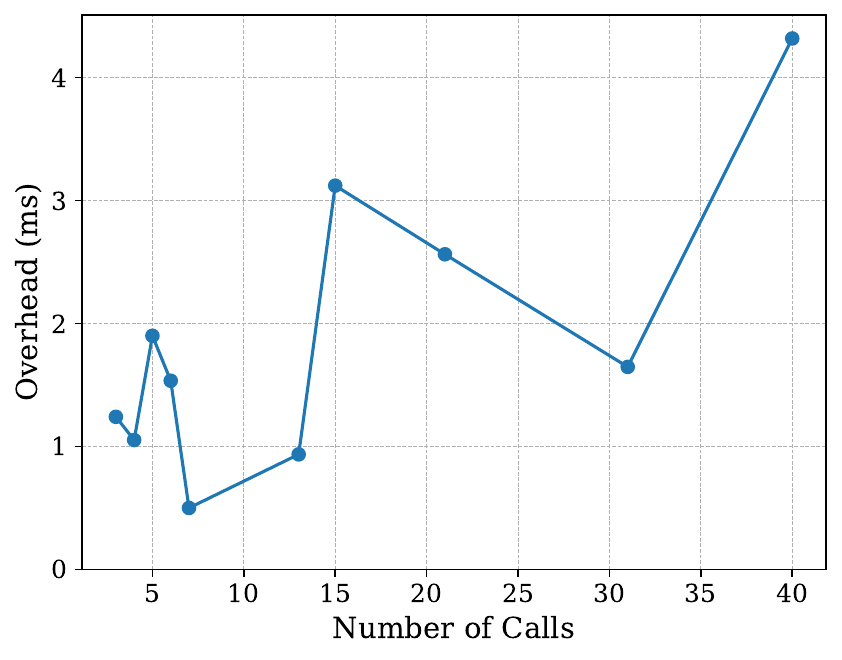}
        \caption{Overhead vs $\le$40 Calls}
        \label{fig:scalezoom}
    \end{subfigure}
    \caption{Latency overhead vs topology scale, measured as number of API calls.}
    \label{fig:scalability}
\end{figure}

\newpar{Overhead versus topology scale.}
To assess \safenom monitor's scalability, we measure the increase in latency overhead with the scale of the application's topology, which we measure as the number of API calls in an execution trace. To measure this, we synthetically generate applications with execution traces ranging from $3$ to $1365$ API calls by considering different application topology (or call tree) shapes---all combinations of depths from $2$ to $5$ and fan-outs from $1$ to $4$. \Cref{fig:scaleall} shows that the average latency overhead in milliseconds (on the y-axis) linearly increases with the number of calls in the trace (plotted on the x-axis). Both axes in the plot use a logarithmic scale.  As per Alibaba's study \cite{alibaba}, the common-case execution traces have fewer than $31$ API calls. As shown in \Cref{fig:scalezoom}, which zooms into the latency overhead plot for traces of length at most $40$ API calls, the latency overhead for common case traces is less than $4$ms.

\begin{wrapfigure}{r}{0.40\textwidth}
    \vspace{-1.5ex}
    \centering
    \includegraphics[height=3.3cm,width=\linewidth]{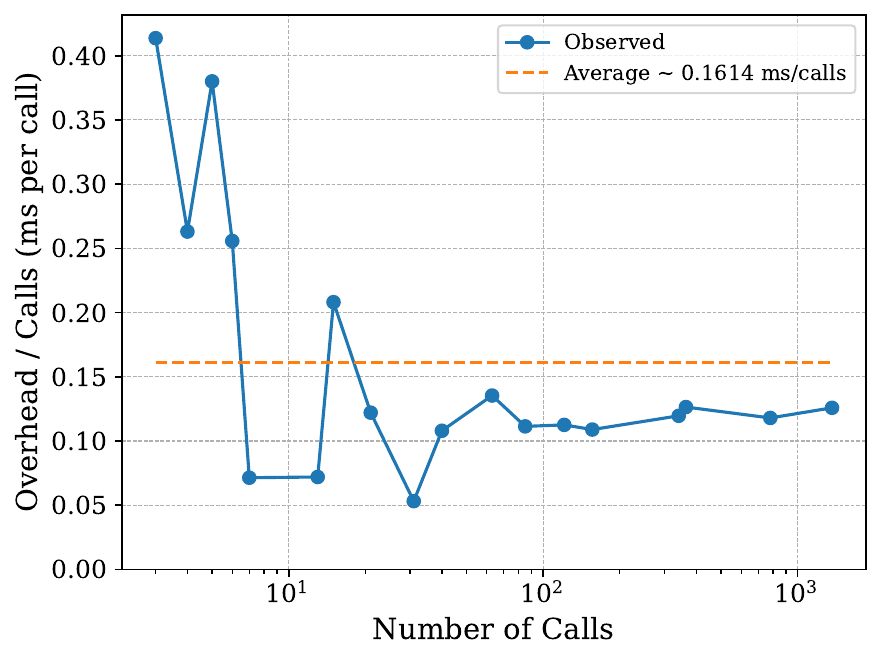}
    \caption{Ratio of overhead to calls.}
    \label{fig:overheadfactor}
    \vspace{-2ex}
\end{wrapfigure}
We also plot the per-hop monitoring overhead (latency overhead divided by number of API calls in the trace) in milliseconds on the y-axis of \cref{fig:overheadfactor} and the number of API calls (on log scale) on the x-axis. Each hop incurs an average $0.16$ ms overhead. \textbf{\textit{In summary, \safenom monitor is scalable with the common-case latency overhead under $\sim 4$ ms and a linear increase in the overhead with the length of the trace, adding 0.16ms of average per-hop overhead.}} 

\section{Related Work}
A large body of work studies automata whose input symbols are drawn from an infinite or a very large domain. Representative models include register automata, finite-memory automata, fresh-register automata, pebble automata, class-memory automata, history-dependent automata, and nominal automata \cite{ra,freshra,pebble,hda,na1,infdatasurvey}. These models differ in how they store names, compare them, or enforce freshness. These models are relevant to \safenom because our policies must relate names at different positions for equality patterns. However, \safenom also needs a language model with first-class support for binding, scope, and alpha-renaming notions, rather than only a part of the automata implementation. Below we survey some automata models with register-style implementation and nominal formalism like our model, followed by work on runtime monitoring of microservice applications.  

\newpar{Register-based Automaton and languages over infinite alphabet}
Register-based automata vary in how they handle the memory and here we describe some of the key details. 
Finite memory automaton (FMA) \cite{fma} and register automaton (RA)\cite{ra,symra} extend a finite-state automaton with a fixed finite set of registers to  store and compare data values in a flat word with values, and depending on the model update registers. The maximum number of values that can be simultaneously tracked is statically fixed in the automaton definition. These models can express equality and inequality relationships between values in the flat trace. Some variants can test local freshness, \textit{i.e.}, the current value is not in one of the registers, while others like fresh-register automaton \cite{freshra} can test a form of global freshness, where the current value must be fresh with respect to the entire run.

These models offer a promising automata machinery for data-aware monitoring, but they do not make  lexical scope  and alpha-renaming a primitive parts of the language semantics. One could extend these models to add an ad-hoc register allocation/deallocation discipline to store and compare values within an active scope. However, then the scope information gets baked into the automata machinery. In contrast, \safenom makes binders and scope explicit in the policy language. Our automaton still has register-based operational semantics.

\newpar{Nominal Automata}
Nominal automata study languages over an infinite alphabet through the lens of symmetry \cite{symaut,10.1007/s001650200016,rbe,nka2}. In the standard nominal setting, the concrete names are not important in themselves; the only aspect that matters is how names are compared, reused, and required to be fresh. Formally, states and transitions are equivariant under permutation of names, so the automaton can be orbit-finite---the automaton has finitely many states up to consistent renaming--despite names ranging over an infinite (or large) set of names. The nominal view is the right starting point for \safenom as our policies should be invariant under consistent renaming of API values. The policy should not depend on the concrete values carried by API calls and returns, but only on the values' equality, inequality, freshness, and reuse patterns. However, the general nominal automaton model is too abstract to be directly operationalized into a deployable monitor in the microservice setting. Instead, we build on the register-style nominal automaton model by \citet{kurznomautomata,kurz1,kurz2} rather than a fully general nominal automaton.

\newpar{Binder-based nominal automata}
The closest technical foundation for \safenom is the binder-based nominal automaton model of \citet{kurznomautomata}. This work builds on history-dependent automata \cite{hda}. History-dependent automata already provide a means to allocate names, compare, rename, and forget them; but they recognize flat words. However, in our setting,  input words have explicit binders. Many binder-based automata models are designed to characterize the language expressiveness and often allow non-deterministic choice about name generation, matching, and binding \cite{kurz2,nkleene,nondetfma}. Our online monitor should deterministically update its state by observing just one symbol at a time; it cannot backtrack its transitions. So we go with the binder-based automata model by \kurz, which satisfies our need for determinism.

\newpar{Runtime monitoring of data-aware properties} 
There is a plethora of runtime verification work for systems with parametric events \cite{javamop,ruler,qea,logscope}, register automata-based (invasive) monitoring of Java programs \cite{regautrv}, and more recently stream runtime verification \cite{streamrv}. These systems focus on parametrized monitoring, but \safenom focuses  on policies with primitive support for scope, binding, and freshness. One data-aware monitoring system, BeepBeep \cite{rvwebinterface,rvmessagecontracts,rvdataaware} is close in spirit to \safenom . It is an LTL-based runtime enforcement mechanism for safety properties over sequences of web service calls and first-order quantification over data. Unlike \safenom , it does not support properties that require freshness and scoping constructs. Also, BeepBeep compiles to a more complicated B\"{u}chi automata-based monitor in comparison to lightweight nominal automata. 

\newpar{Monitoring microservice safety properties}
Several recent works generalize microservice safety properties \cite{safetree,grewalhotnets,copper} beyond pair-wise properties and offer a service mesh based monitoring mechanism. SafeTree \cite{safetree} shows that API-level monitoring is a practical way to enforce safety properties for blackbox microservice applications. Local monitors deployed outside each service running atop service mesh layer can observe all the API calls/returns into the service and efficiently enforce the properties using a  relevant automaton model. \safenom also uses this service mesh-based deployment for its monitor to meet its monitor's non-invasive and blackbox deployment requirements. 

However, SafeTree and related systems abstract away the API parameters and can only enforce properties over an execution's control-flow, like which API may call which API, which
API must happen before another API, or which service call tree shapes are permitted. Such policies are useful, but they are insufficient for the data-aware safety properties described in this paper. \safenom addresses this orthogonal dimension of data-aware properties. It retains the blackbox API-level view of 
microservice execution, but enriches the
events with logical names that represent parameter values. \safenom policies can then specify
both the order of API events and the scoped equality or inequality relationships among
their parameters and responses, and enforce these properties using a nominal automaton.

\section{Conclusion}
We present \safenom , a nominal language-based policy language to specify data-aware microservice safety properties. \safenom policies are enforced in a blackbox manner using a nominal automaton-based runtime monitor that runs efficiently without any invasive changes to the service implementations. \safenom currently assumes that API calls are sequentially invoked. We see a future possibility of  extending \safenom to an asynchronous setting, where an API call sends data in parallel to other services.

\bibliography{nominal}
\newpage
\ifappendix
\appendix
\fi
\end{document}